\documentclass[a4paper,notitlepage,12pt]{jedm}

\usepackage[table]{xcolor}
\usepackage{url}
\usepackage{amsmath}
\usepackage[bb = boondox]{mathalpha}
\usepackage{multicol}
\usepackage{tikz}
\usetikzlibrary{positioning}
\usetikzlibrary{backgrounds}
\usepackage{multirow}
\usepackage{graphicx}
\usepackage{bbm}
\usepackage{algorithm}
\usepackage{algpseudocode}
\usepackage{soul}
\usepackage{xcolor}
\usepackage{color}
\usepackage{amsmath}
\usepackage{amsfonts}
\usepackage{wrapfig}
\usepackage{bm}
\usepackage{booktabs}
\usepackage{algpseudocode}
\usepackage{enumitem}
\usepackage{hyperref}
\let\bibhang\relax
\usepackage{natbib}

\newcommand{\redacted}{frederikbaucks}

\begin{document}

\title{The Course Difficulty Analysis Cookbook}
\date{}

\author{{\large Frederik Baucks}\\Ruhr-Universität Bochum\\frederik.baucks@ini.rub.de 
\and {\large Robin Schmucker}\\Carnegie Mellon University\\rschmuck@cs.cmu.edu 
\and
{\large Laurenz Wiskott}\\Ruhr-Universität Bochum\\laurenz.wiskott@ini.rub.de}

\maketitle

\begin{abstract}
Curriculum analytics (CA) studies curriculum structure and student data to ensure the quality of educational programs. An essential aspect is studying course properties, which involves assigning each course a representative difficulty value. This is critical for several aspects of CA, such as quality control (e.g., monitoring variations over time), course comparisons (e.g., course articulation), and course recommendation (e.g., student advising). Measurement of course difficulty is a nuanced problem that requires careful consideration of multiple key factors: First, when difficulty measures are sensitive to the performance level of enrolled students, it can bias interpretations by overlooking diversity in student performance. By assessing difficulty independently of enrolled students' performances, we can reduce the risk of bias and enable fair, representative assessments of course challenges. Second, from a measurement theoretic perspective, the measurement must be reliable and valid to provide a robust basis for subsequent analyses. Third, difficulty measures should be nuanced and account for covariates, such as the characteristics of individual students within a diverse populations (e.g., transfer status, dropout graduation status). In recent years, various notions of difficulty have been proposed. This paper provides the first comprehensive review and comparison of existing approaches for assessing course difficulty based on grade point averages and latent trait modeling. It further offers a hands-on tutorial offering guidance on model selection, assumption checking, and practical CA applications. These applications include monitoring course difficulty trends over time and detecting courses with disparate outcomes between distinct groups of students (e.g., dropouts vs.\ graduates), ultimately aiming to promote high-quality, fair, and equitable learning experiences. To support further research and application, we provide an open-source software package named 'Course Difficulty Estimation' (CDE)\footnote{\href{https://github.com/\redacted/course-difficulty-estimation}{https://github.com/\redacted/course-difficulty-estimation}} and artificial datasets with an implementation of methods, including documentation facilitating reproducibility of analyses and method adoption. 

{\parindent0pt
\textbf{Keywords:} course difficulty, grade point average, item response theory, additive model, tutorial
}

\end{abstract}

\section{Introduction} 
    % Understanding how specific course characteristics, such as difficulty, influence student outcomes remains a relatively unexplored area.
    While some research has explored how specific course characteristics, such as difficulty, can affect student outcomes, our understanding these relationships is incomplete and warrants further research.
    Curriculum Analytics (CA) addresses this gap by focusing on course-specific factors that impact student success \citep{romero2019edm}. The objectives of CA include ensuring alignment between course content and learning objectives, optimizing prerequisite structures, and establishing course quality measures. By providing insight into these areas, CA provides valuable guidance for curriculum development and continuous improvement. As a prerequisite, curriculum data provides the details: course content, structure, and assessment data. It is the "what" we teach and the "when" and "how" we assess it, ultimately resulting in the generation of grades. Using this data, CA methods seek to identify and understand the factors that contribute to student outcomes, including grades (e.g., \citealt{baucks21mitigating,baucks24lak}), dropout (e.g., \citealt{salazar2021curricular,aina2022determinants}), and time to degree  (e.g., \citealt{molontay2020characterizing,baucks2022simulating}). Methods include process mining \citep{wagner2022combined}, simulation \citep{saltzman2012simulating,molontay2020characterizing}, and curriculum-based prediction \citep{backenkohler2018data}, e.g. with Bayesian belief networks \citep{slim2014predicting}. These CA methods turn the curriculum-related data into insights. Finally, stakeholders - from students to policymakers - rely on these insights to make informed decisions that enhance the curriculum's relevance and effectiveness. Together, these elements form a continuous improvement cycle, making CA a critical part of educational development \citep{hilliger2022lessons}. 
    
    Besides helping us understand the effects of deliberate decisions within educational institutions, CA also gauges the effects of unanticipated factors, such as external influences (e.g., the COVID-19 pandemic) or internal changes (e.g., teachers exploring new instructional methods). As a consequence, these factors might influence student outcomes. Identifying these causalities in student outcomes is difficult because multiple factors can act simultaneously (e.g., teachers, student population, course content). However, neglecting these nuances can yield misleading insights. Student grade point averages (GPA) and course difficulty, as often measured by pass rates, illustrate this. Course difficulty is an essential statistic for measuring curriculum effectiveness and quality; for example, if a course’s difficulty deviates significantly from the average, it may block students' further progress (high difficulty) or indicate redundancies between courses (low difficulty). Given the interdependencies between students and courses - such as how pass rates are affected by student GPA - it is critical to disentangle the factors within the learning environment that affect course difficulty. Stakeholders such as student advisors and program planners need trustworthy course difficulty estimates to keep the curriculum effective. Student advisors assume constant course difficulty over time and need to be aware of difficulty variations to provide consistent academic advice \citep{baucks2024book}, and program planners might use course difficulty to identify and address potential bottlenecks in the curriculum \citep{saltzman2012simulating}. 

    In recent years, multiple approaches for quantifying course difficulty have been proposed. The state-of-the-art approaches utilize various statistical and machine learning techniques applied to course grade data. Initial approaches for quantifying course difficulty take student grades and compute mean course grades and GPAs to measure difficulty and performance (e.g., \citealt{molontay2020characterizing,saltzman2012simulating}). However, the difficulty of a course often depends on factors such as the performance of the students enrolled in the course or the teacher teaching it. One limitation of many approaches is that they do not explicitly decouple these factors fully. If individual factors are not decoupled, they can confound difficulty estimates, leading to biased interpretations of course difficulty \citep{baucks21mitigating}. For example, a difficult course may seem less difficult because a particularly strong cohort of students took it. Researchers have proposed adjusted course difficulties using one centering approach and two latent trait modeling approaches to decouple course and student factors to address these shortcomings:
    % ---- Centering approaches
    Firstly, \textit{centering-based difficulty adjustments}, as described by \cite{ochoa2016simple}, attempt to account for the performance of enrolled students in course pass rates by centering all grades in a course on their corresponding student GPAs. Assuming that the GPA is sufficiently representative of a student's performance, \cite{mendez2014curricular} suggests that centering can lead to valid estimates of difficulty that correlate with students' perceived difficulty.
    % ---- IRT
    Secondly, \textit{Item Response Theory} (IRT) was initially developed for high-stakes assessment and uses statistical techniques to measure latent traits of test takers and the difficulty of test items \citep{Ayala2013:Theory}. IRT models student's responses to multiple test items (e.g., multiple-choice questions) by assuming a student's response to a given item can be explained by a probabilistic relationship between the student's trait and the item's difficulty. In CA research, recent studies have successfully applied IRT-based methods, leading to course measures that account for variations in ability levels among enrolled students (e.g., \citealt{Bacci2017:Evaluation,baucks24lak}). 
    % ---- AGM
    Thirdly, in GPA adjustment research, \textit{additive grade point models} (AGM) have been developed on continuous data to model course grades linearly estimating latent traits for individual students and courses. In CA, AGMs measure course difficulty, adjusting for student performance factors (e.g., \citealt{caulkins1996adjusting,baucks21mitigating}).

    This paper critically examines the strengths of difficulty models (i.e., IRT models and AGMs) and the limitations of unadjusted heuristic approaches in course difficulty assessment (e.g., student GPA, course pass rates). While unadjusted pass rates remain a commonly used measure despite their known weaknesses (e.g., \citealt{srivastava2024curriculum}), we offer practical guidance on using adjusted models, such as latent variable estimation and centering, to improve reliability and validity. These models introduce complexity and require rigorous statistical validation. Hence, we outline a streamlined approach to ensure model reliability and usability. This tutorial provides readers with a framework for selecting and applying the best difficulty estimation method for their personal CA needs.
    
    Consequently, this paper presents a tutorial (including a hands-on tutorial) for modeling course difficulty based on student grade data. In a comprehensive methodology, we show which difficulty estimation methods best fit the course grade data for different grade types (e.g., binary and continuous), assess model fit and assumptions, and highlight their applications on real data sets. These applications show that estimates of course difficulty can answer important CA-related research questions that heuristics can not (e.g., has a course gotten more or less difficult due to a change in course characteristics or student population, and what is the impact of a teacher change?). 
    In this regard, this work provides researchers and practitioners with hands-on guidance for estimating course difficulty, thus providing a solid foundation for assessing curriculum quality.
    The main emphasis of this work lies in helping researchers and practitioners leverage these techniques to answer their CA-related questions by provide guidance for choosing a suitable model, verifying underlying model assumptions, and assessing measurement properties. Our contributions include:

    \begin{itemize}   
        \item \textit{Comparison of Difficulty Estimation Methods:} We provide an overview of methods for modeling course difficulty from course grades and compare them in various simulated data settings. We consider two main model types: firstly, heuristic models and their centering-based versions, and secondly, latent variable models, including item response theory and additive grade point models, each determining course difficulty via statistical inference. Based on the grade type in the data (e.g., binary or continuous course grades), we provide guidance on which model type to choose.
        \item \textit{Guidance to check Assumptions:} Without checking the assumptions of a model, its application can lead to misleading insights. We provide a detailed overview of the model assumptions. In particular, we present a methodological pipeline for testing these assumptions. We extend the standard literature assumption tests to include missing data, a common occurrence in CA-related course grade datasets.  Although the assumptions are the same for all three modeling approaches, their verification can differ depending on the data (e.g., binary or continuous). Furthermore, we assess the robustness of the proposed experimental design using simulations gauging the influence of different missing value proportions.
        \item \textit{Assessing Measurement Properties:} Because course difficulty values inform the de\-cision-making processes of various stakeholders, we need to ensure the validity and reliability of the difficulty measurement process. Validity refers to the accuracy of a model in measuring what it is intended to measure, ensuring that difficulty estimates truly represent course difficulty as conceptualized. Reliability refers to the consistency of the estimates, meaning reliable models produce similar results when repeated on different samples. Therefore, assessing these two properties is essential for reproducibility and accurate interpretation. After going through the assumption-checking pipeline and fitting models, we build a separate set of experiments to assess the reliability and validity of the model parameters. %These checks lead to robust parameters that represent the concept of course difficulty. 
        \item \textit{Case Study on German University Data Set:} We illustrate the utility of the proposed CA pipeline by applying it to real data from a German university. Using data consisting of grades of nearly 2000 students in about 30 courses spread over nine years in two majors, we walk the reader through the individual steps of the methodology and showcase how it can be used to address various CA questions. We first verify that modeling the difficulty of the courses with latent models satisfies the corresponding assumptions of the models. We then generate various insights for stakeholders -- including student advisors, curriculum policymakers, and teachers -- by quantifying the impact of external events and analyzing differences between student cohorts.
        \item \textit{Baseline Simulated Data:} 
        We provide simulated data that generate upper bounds that complement existing lower bounds identified in the literature as critical values (e.g., minimum correlation values) necessary for the assumption checking and measurement property experiments. This approach allows us to evaluate the proposed methodology's performance and quality comprehensively.
    \end{itemize}
    The paper is structured as follows: After an overview of related work, a \hyperref[sec:hands_on]{\textit{hands-on tutorial}} introduces users to the practical aspects of modeling course difficulty using our open-source package 'Course Difficulty Estimation' (CDE). The section is designed to provide an accessible entry point so that users can start exploring CA questions of their personal interest. The CDE package facilates the correct application of the methodologies by automating experiments as well as assumption checks, supporting users in conducting rigorous analyses. For inclined readers, the \hyperref[sec:theory]{\textit{methodological tutorial}} delves into the detailed modeling pipeline, covering the methodological nuances of course difficulty estimation. The subsequent \hyperref[sec:applications]{\textit{case study}} section illustrates real-world use cases, highlighting how the analysis pipeline can be applied to answer questions in various educational contexts. Finally, the \hyperref[sec:discussion]{\textit{discussion}} section reflects on methodological limitations, future work, and broader implications.

\section{Related Work}

    % Curriculum Analytics
    Curriculum analytics (CA) evaluates educational program structure and effectiveness for continuous refinements \citep{hilliger2020design}. An effective curriculum consistently challenges students with relevant learning content \citep{kumar2022responsible} while ensuring fair assessment across students \citep{luke2013curriculum} (e.g., from different cohorts). Besides research on content relevance (e.g., alignment with employers' expectations), most quantitative research in CA focuses on process mining (e.g., \citealt{brown2018taken,wagner2022combined,martinez2023evaluation}) and simulating students' paths through a curriculum (e.g., \citealt{molontay2020characterizing,mceneaney2022curriculum,saltzman2012simulating} or predicting students' outcomes based on the structure of the curriculum (e.g., \citealt{slim2014employing,backenkohler2018data,pardos2020university}). Process mining extracts, analyzes, and models the sequences of interactions that students have with diverse educational components, such as courses, assignments, or learning activities. Process mining helps us understand the pathways students take to navigate through a curriculum and identify courses with unintended properties (e.g., bottleneck courses, which hinder progress if they are failed because they are a prerequisite for other courses). When processes change or are intended to change (e.g., changing recommended course order), simulation methods can be used to predict the changes' impact on student experiences (e.g., course outcomes and graduation time). Finally, predictive models focus on estimating students' future outcomes and are used to finetune simulations or provide personalized recommendations for students' curricular pathways (e.g., student advising).

    % Course difficulty
    Given these methods, one area of particular interest is assessing course difficulty, which is a crucial factor in CA questions \citep{ochoa2016simple}. Course difficulty modeling can help to estimate and promote desired assessment properties, including equity (e.g., between students from diverse backgrounds) and fairness (e.g., between similar students in different cohorts) \citep{baucks24lak}. All three CA method categories, process mining, simulation, and prediction, typically make limiting assumptions about course difficulties by homogenizing the student population or assuming constant course properties over time. Process mining typically assumes that course difficulty is constant over time, a violation of which is known as the phenomenon of concept drift \citep{bogarin2018survey}. As a consequence, simulations can also suffer from this. Predictive models usually assume that course difficulty is independent and identically distributed (iid), which is at risk if courses are aggregated over time \citep{baucks24lak}. Stakeholders relying on the insights generated by CA methods can carry the simplified difficulty assumptions further into decision-making processes. For example, articulation officers need to assess or assume course difficulty to align with standardized benchmarks to facilitate credit transfer \citep{pardos2019data}. Program planners might use course difficulty to identify courses in the curriculum that block students \citep{saltzman2012simulating} and simulate graduation time changes after adjustment \citep{molontay2020characterizing,baucks2022simulating}. Thus, course difficulty is a central concept in research and practice. However, the traditional methods of assessing course difficulty rely on simple grade averages or medians (e.g., \citealt{ochoa2016simple,mendez2014curricular,srivastava2024curriculum}), which can be confounded by the performance of enrolled students and other factors inducing variation \citep{boeve2019natural}, e.g., teachers, and students' economic background. Studies (e.g., \citealt{lei2001alternatives,baucks21mitigating}) have highlighted these limitations and identified reliability issues and better prediction validation after adjustments (e.g., \citealt{caulkins1996adjusting,baucks24lak}), advocating for more sophisticated statistical techniques. These include centering approaches, item response theory-based  (IRT) methods \citep{baucks24lak}, and linear additive grade point models (AGM) \citep{baucks21mitigating}.

    % Centering intro 
    Centering approaches to course difficulty estimation use the grades in a course and subtract the GPAs of enrolled students of each corresponding grade. These approaches attempt to reduce the influence of student performance on the course difficulty estimate. The use of such transformations originated in research on GPA adjustment. For example, \cite{caulkins1996adjusting} have adjusted students' GPAs at a US college to mitigate divergent grading standards in different courses of the same major. \cite{johnson2006grade} have used centering to compare grading systems in different majors and have concluded that students' course choices depend on the grading practices in the courses available for selection. In recent years, these estimates have also been examined in the context of course difficulty in CA \citep{ochoa2016simple,mendez2014curricular}. Here, average \textit{course} grades are transformed instead of average \textit{student} grades, resulting in course difficulty estimates. Research shows correlations between the estimated course difficulties and perceived difficulties as captured by student questionnaires \citep{mendez2014curricular}. However, the grades students received in the course to be rated might bias students' personal perception of course difficulties \citep{wang2021debiasing}.
    
    % IRT intro
    IRT models the relationship between latent traits (such as student abilities) and their \textit{binary} performance on assessment items, providing insights into item characteristics, including difficulty. The methodology is commonly employed in educational research to model student abilities and item difficulties in the context of high-stakes testing \citep{Ayala2013:Theory, lord1980applications}. 
    IRT methodologies are foundational in modern item difficulty research, e.g., in the OECD PISA studies \citep{pisa2024technical}. It has also been adapted to different contexts than standardized testing, for example, GPA adjustment in the university context \citep{caulkins1996adjusting,hansen2019estimating}, where the items and their responses are replaced by courses and their 'pass/fail' grades. These adjustment studies, in particular, highlight the importance of course factors and variance influencing students' performance. Similarly, recent advances explored the use of IRT for analyzing higher education data, focusing on assessing course-specific properties, in particular course difficulties \citep{bacci2015classification,haas2023bayesian,baucks24lak}.

    % AGM intro
    AGMs model \textit{continuous} student grades using linear but independent factors, e.g., each student and course is assigned a factor, which is then identified as student performance and course difficulty. AGMs offer a flexible approach to handling confounding variables (e.g., student's learning rates and course-teacher dependencies) in educational data \citep{boeve2019natural} since AGMs can accomodate more factors such as learning rates \citep{koedinger2023astonishing}. Research has shown how these models can isolate the effect of course content from student performance factors (e.g., \citealt{beenstock2018decomposing,baucks21mitigating}). These efforts underscore the importance of addressing confounding factors to obtain reliable course difficulty estimates.
        
    % Models need assumption, reliability, validity checks and this implies the problem we are tackling using the cookbook receipt.
    While difficulty estimates by centering are easy to implement, IRT models and AGMs are statistically more sophisticated in modeling course difficulty and offer frameworks for exploring nuanced CA questions. The effectiveness of all three models heavily relies on checking underlying assumptions and ensuring the models' reliability and validity. First, testing model assumptions is essential to achieving robust parameter estimates and results, yet this step is often overlooked \citep{bergner2017measurement}. This may be because these model assumptions are often difficult to test with real-world data, e.g., due to missing data, as they require nuanced statistical considerations. However, neglecting these checks can lead to inaccurate estimates of difficulty, undermining the utility of the model in practical applications \citep{baucks21mitigating}. Secondly, the concepts of measurement validity and reliability are critical. Validity refers to the accuracy of a model in measuring what it is intended to measure, while reliability pertains to the consistency of the model's estimates. For example, in the context of course difficulty estimates, a valid model accurately reflects the actual difficulty of courses, and a reliable model provides consistent difficulty estimates across different cohorts. Failing to ensure these aspects can result in significant issues: unreliable models might suggest changes to a curriculum based on inconsistent data, and invalid models might mislead stakeholders about the actual difficulty of courses, impacting decisions like academic advising and curriculum planning. These challenges are particularly pronounced due to the complexity and variety of educational data, making the rigorous testing of assumptions and measuring reliability and validity complex. Handling different data types (binary, categorical, continuous) and dealing with missing values add another layer of complexity. 

    % Contributions 
    This work provides researchers and practitioners with a practical hands-on tutorial for implementing key models in curriculum analysis, focusing on centering, IRT, and AGM. The hands-on tutorial guides users through the model application process. It provides a structured foundation for conducting accurate assessments of course difficulty and an overview of addressable CA questions. The subsequent methodological tutorial explores essential considerations for model selection, missing data handling, and assumption checking, equipping users with the knowledge to make robust methodological choices for their needs. Finally, our case study applies the models to assess the impact of external events on course difficulty, differences between dropouts and graduates, and differences between student cohorts.

\section{Hands-On Tutorial}\label{sec:hands_on}

    In this section, we present the hands-on tutorial providing readers with a high-level overview of the methodology and how it relates to the 'course difficulty estimation' (CDE) package. The CDE package is available in an open-access GitHub repository\footnote{\href{https://github.com/\redacted/course-difficulty-estimation}{https://github.com/\redacted/course-difficulty-estimation}}, in which we also provide a quick start tutorial using simulated data. Lastly, we list examples of research questions that can be answered with our CDE package. Later in the paper, in the Case Study section (Section \ref{sec:applications}), we demonstrate how our methodology addresses these questions using real data.

\subsection{High-Level Methodology and CDE package}
    CDE combines statistical modeling, assumption checking, reliability checks, and validation checks to assess course difficulty and student performance. In addition, it can account for group differences in course difficulty and thus can inform various applications, such as designing tailored support for individual students.
    At a high level, we rely on the following:
    \begin{itemize}
        \item  \textit{Latent Trait Models:} These models estimate an underlying "difficulty" parameter for each course and a "performance trait" parameter for each student derived from course grade data. By fitting a latent trait model (e.g., Item Response Theory model), our method captures how students of different performance traits interact with courses of different difficulty levels, producing interpretable, robust estimates of these metrics.
        \item  \textit{Regression-based adjustment for group differences:} To assess potential differences in perceived difficulty between groups of students, the method uses a regression analysis called differential course function (DCF) to compare performance across groups (e.g., demographic categories). We can estimate group-specific effects on course performance independently of individual student performance traits and the courses' global difficulty. This isolates group-specific effects, allowing users to identify potential disparities related to the groups, e.g., caused by language barriers.
    \end{itemize}
    \begin{figure}
        \centering
        \includegraphics[width=0.85\linewidth]{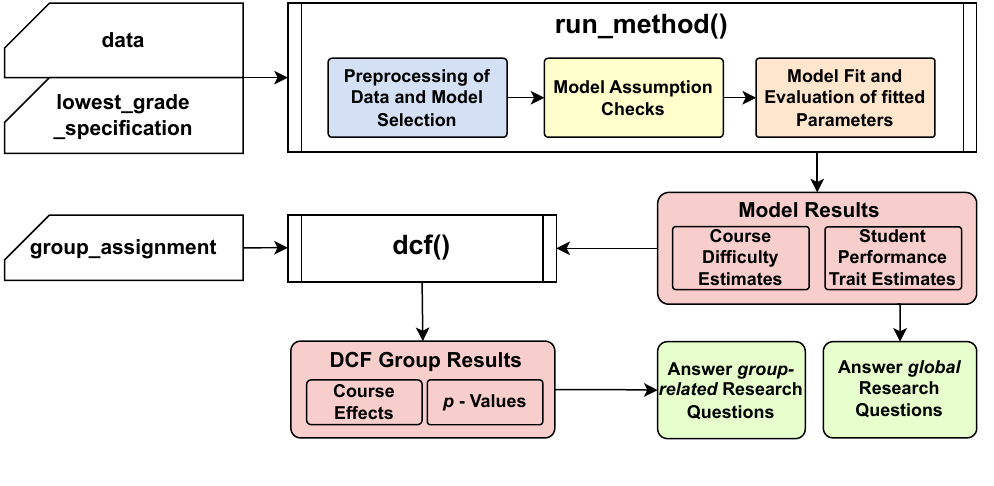}
        \caption{High-level overview of the methodology implemented by the CDE package. It contains two possible paths, depending on the research objective. If only course difficulty is to be estimated use the \texttt{run\_method()} function. If group-specific course difficulties are also to be estimated, the model results and \texttt{group\_assignment} data are used to call the \texttt{dcf()} function, which returns the group-specific course difficulties. Please consult the methodological tutorial using the same color coding for a more detailed discussion.}
        \label{fig:high_level_meth}
    \end{figure}

    In the following, we will go through the steps necessary to apply our CDE package. Figure \ref{fig:high_level_meth} shows a high-level methodology overview. 
    We introduce the functions \texttt{run\_method()} and \verb|dcf()|. Firstly, \verb|run_method()| receives "data" and a "lowest\_grade\_specification" to fit latent trait models. The "data" includes student grades. The "lowest\_grade\_specification" specifies what grades represent high achievements. Depending on the type of grades in the data (e.g., binary or continuous), a suitable model class is chosen (blue box). Then, the class assumptions are checked (yellow box), the model is fitted, and its fit is evaluated (orange box). If both latter two (yellow box and orange box) are sufficient, the method returns course difficulty estimates and student performance trait estimates. Otherwise, the respective check is flagged, and the user should consult the corresponding experiments in the methodological tutorial to address the issue (Section \ref{sec:theory}).
    Secondly, \verb|dcf()| receives group assignments defined by the user and the returned model results of \verb|run_method()| to fit group differences in course difficulty using regression-based methods. The results of either or both methods can be used to answer related research questions such as the ones outlined in Table~\ref{tab:research_questions} (green boxes). The methodological tutorial in Section \ref{sec:theory} provides detailed breakdown of each step. 
    Figure \ref{fig:high_level_meth} and the methodological tutorial use the same color scheme for better orientation.

    \subsubsection{Data Preparation}
        The CDE package expects course grade data as input. To calculate course difficulties, the user must specify a course response matrix \verb|data| containing the students' grades. This is constructed in a pandas~\citep{pandas} DataFrame as follows: The rows represent the students; each student must have a unique identifier corresponding to the DataFrame's index. The DataFrame columns represent the courses, and the column names define the course names. The entries in the DataFrame are the students' course grades (e.g., binary grades or percentage grades). The user must pre-process the course grades so that non-missing grades contain only numerical values. The CDE package automatically handles missing values in NumPy's not a number representation \verb|numpy.nan|~\citep{harris2020array}.
        \begin{verbatim}
data = 
     course_A  course_B  course_C  course_D  course_E    ...
 s_1      nan         1         1         0         1    ...
 s_2        0         1       nan       nan         1    ...
 s_3      nan         1         1       nan         1    ...
 s_4        0         0         1       nan         1    ...
 ...      ...       ...       ...       ...       ...    ...
        \end{verbatim}
        To estimate difficulties according to their scale (here, it is a binary scale), the lowest grade and order of grades need to be specified. For example, the US grading system typically ranges from $0$ to $4$ with $4$ being the best grade--corresponding to function parameters $0$ and 'ascending'.
        \begin{verbatim}
lowest_grade_specification = (0, 'ascending')
        \end{verbatim}
        In contrast, the German grading system typically ranges from $5$ to $1$ with $1$ being the best grade--corresponding to function parameters $5$ and 'descending'.
        \begin{verbatim}
lowest_grade_specification = (5, 'descending')
        \end{verbatim}
        Then, the model can estimate the course difficulty and student performance trait estimates according to the grade types (here in $\{$\verb|0|$,$\verb|1|$\}$) in the matrix. A Jupiter notebook with simulated data in the GitHub repository details the process and provides a reference implementation.

    \subsubsection{Implemented Estimation Functions}
        Once the DataFrame is created, the \verb|run_method()| function can be called. From there on, the repository automatically checks the model assumptions of the models used, performs model fitting, ensures the robustness and validity of the estimates, and finally outputs the course and student estimates. 

        \begin{verbatim}
Specify:
    data, lowest_grade_specification
            
Call:
    model = run_method(data,
                       lowest_grade_specification)
    course_estimates = model.course_est
    student_estimates = model.student_est
        \end{verbatim}

        In many settings, it is important to assess systematic differences between distinct student groups. For example to answer whether a specific course disadvantages certain individuals (e.g., transfer students). The CDE package allows the user to assess these differences. The corresponding function \verb|dcf()| fits a regression model that assesses the difference between two groups of students independently of \verb|student_estimates| and \verb|course_estimates|. This ensures that the fitted differences between the groups are not due to general performance differences. In addition, \verb|dcf()| returns a $p$-value indicating whether the group difference \verb|course_effect| is significantly different from zero. To fit the regression model, the user needs to specify the \verb|course_name| of the respective course, which needs to match the column name of that course in \verb|data|. In addition, \verb|group_assignment| needs to be specified. This is a pandas DataFrame with a column consisting of student names and a column indicating the group assignment of each student using the values $-1$ and $1$. Note that when performing multiple tests (e.g., for all courses in the dataset), it is necessary to adjust the significance level to control the false discovery rate (FDR). While a threshold of $\alpha = 0.05$ is typically used for a single test, in the case of multiple tests we recommend applying the Benjamini–Hochberg correction \citep{Baucks24Gaining_LAS}.

        \begin{verbatim}
Specify: 
    course_name, group_assignment
            
Call:
    course_effect, p_value = dcf(data, 
                                 student_est, 
                                 course_name, 
                                 group_assignment)
        \end{verbatim}

    \subsubsection{Assumption Checks and Measurement Properties}

    Course difficulty models make theoretical assumptions that must be verified when they are applied to real-world data. The statistical tests required to evaluate these assumptions are discussed in detail in the methodological tutorial (Section~\ref{sec:theory}). The methods implemented in the CDE package automate these tests to check whether the real-world data meet the assumptions. A flag is raised if one of the model assumptions is at risk of being violated. In this case, caution is advised, and the user should refer to the methodological tutorial, which outlines directions on how to proceed. Otherwise, the user can continue working with the difficulty estimates to answer research questions of interest. Representative examples of research questions that the analysis pipeline can address are illustrated in Table \ref{tab:research_questions}.

\begin{table}[ht!]
\centering
     \caption{Examples of research questions that can be addressed using the CDE package. The questions are categorized by stakeholder. This list serves for illustrative purposes and is not exhaustive. Questions with references at the end are illustrated with real-world data in Section~\ref{sec:applications}, where the corresponding tables or figures present the case study results.}
     
 \resizebox{\textwidth}{!}{

    \begin{tabular}{|p{0.27\textwidth}|p{1\textwidth}|}
        \hline
        \textbf{Stakeholder} & \textbf{Research Questions} \\ \hline
        Student Advisors \& Academic Support & 
        \begin{itemize}[noitemsep, topsep=0pt, leftmargin=10pt]
            \item Which course combinations exhibit similar average difficulty?
            \item Can we optimize combinations and sequences according to the difficulty?
            \item Are multiple different factors required to succeed in the courses?
            \item How do difficulty patterns across courses predict student workload?
        \end{itemize} \\ \hline
        Accreditation \& Program Planners & 
        \begin{itemize}[noitemsep, topsep=0pt, leftmargin=10pt]
            \item Are assessments fair for students from different cohorts? - \hyperref[tab:dcf_cohorts]{Table \ref*{tab:dcf_cohorts}}
            \item How do course difficulties compare across institutions?
            \item Do external events influence the course difficulties at my university? - \hyperref[sec:monitoring_difficulty]{Figure \ref*{fig:course_diff_over_time}}
        \end{itemize} \\ \hline
        Articulation officers \& Transfer Students& 
        \begin{itemize}[noitemsep, topsep=0pt, leftmargin=10pt]
            \item Are courses equivalent in content also similar in difficulty across different institutions?
            \item What impact do differences in course articulation pairs have on students' academic pathways?
            \item How do course content and course difficulty relate?
        \end{itemize} \\ \hline
        Identifying Needs of Diverse Student Subgroups using DCF & 
        \begin{itemize}[noitemsep, topsep=0pt, leftmargin=10pt]
            \item What impact do tools and services have on the perceived difficulty (e.g., dashboards)?
            \item Can we detect language barriers in courses?
            \item Do courses show implicit biases that impact groups disproportionately? - \hyperref[sec:dcf_result]{Table \ref*{tab:dcf_drop}}
            \item What difficulty patterns exist between students in diverse living conditions, e.g., part-time, parent, and first-generation students?
            \item What courses increase DCF effects between subgroups of students, and is high difficulty related to that?
        \end{itemize} \\ \hline
        Drop-outs and Graduates & 
        \begin{itemize}[noitemsep, topsep=0pt, leftmargin=10pt]
            \item Are there combinations of difficult courses that are related to dropout?
            \item How does course difficulty affect students' transition to consecutive degrees?
            \item How does course difficulty impact students' career path after university?
        \end{itemize} \\ \hline
        Students' Motivation \& Engagement & 
        \begin{itemize}[noitemsep, topsep=0pt, leftmargin=10pt]
            \item How does difficulty relate to the motivation of students?
            \item Do difficulty outliers affect engagement, e.g., courses that are too difficult?
            \item Can difficulty adjustment change the engagement of students?
        \end{itemize} \\ \hline
        \end{tabular}
        }
        
    \label{tab:research_questions}
\end{table}

\subsection{Overview of Research Questions and Applications}

    To demonstrate the utility of the analyses pipeline, we present research questions that can be addressed using the methodology in Table \ref{tab:research_questions}. Overall, these questions can be divided into two categories. Questions that rely only on student performance traits and course difficulty and questions that require student grouping. The first solely utilizes course grade data. Here, our CDE package outputs the difficulty estimates of the courses. The second requires assigning students to distinct groups. Then, the groups are compared to each other to compute group-specific difficulty factors. 

    \textbf{Case Study Overview}: Using two real-world data sets, our case study uses the CDE package to address three research questions in Table \ref{tab:research_questions}, highlighted with references. These point to the corresponding results in Section \ref{sec:applications}. 
    The datasets capture multiple years of student grades in computer science (CompSci) and mechanical engineering (MechEng) programs at a German university. The CompSci dataset spans nine years (2013-2021) and documents the exam scores of 1,098 students in 19 required courses, with a passing score of 50 on a scale of 0-100. After data preprocessing to ensure privacy and consistency, such as adding ±5 point noise, including only first-time course exam attempts, and requiring at least five grades per student, the final sample included 664 students. The MechEng dataset covers 2012-2021 and includes grades from 3,059 students in 18 courses, initially recorded on a scale of 5.0 to 1.0. These data were transformed to a 0-100 grade scale to standardize the grading, resulting in a sample of 1,651 students. Both datasets were duplicated and transformed to include continuous scores and binary pass/fail versions. While continuous data maximizes information for modeling, the binary format was created to demonstrate the applicability of CDE to this data format.
    The datasets are detailed in Section \ref{sec::data_sets}.

    \section{Methodological Tutorial} \label{sec:theory}
    
    \subsection{Heuristics and Centered Estimates}
    
    Heuristics are methods that arrive at probable statements or workable solutions with limited knowledge and time, seeking a pragmatic trade-off between effort and accuracy \citep{gigerenzer2011heuristic}. They are a widely used class of metrics that attempt to measure concepts such as student performance and course difficulty, commonly using averages such as a student's grade point average (GPA) and a course's pass rate. The simplest model for measuring course difficulty for a course $c$ is to define its difficulty $\delta_c$ as the pass rate or average grade of a course. Similarly, student performance can be approximated by GPA:
    \begin{align} \label{course_difficulty_pass_rate}
        \delta_c &= \frac{1}{|S_c|}\sum_{s \in  S_c} g_{s,c} &  \text{gpa}_s = \frac{1}{|C_s|}\sum_{c \in  C_s} g_{s,c}
    \end{align}
    where $S_c$ is the set of all students in course $c$, $C_s$ is the set of all courses student $s$ attended, and $g_{s,c}$ is the course grade of student $s$ in course $c$. However, because of their pragmatic focus, heuristics are based on simplifying assumptions, such as the independence of course difficulty from the level of performance of the students enrolled. Recent studies in CA have shown that such assumptions can lead to confounding, and results must be interpreted cautiously to avoid biased interpretations (e.g., \citealt{baucks24lak,baucks21mitigating}). 
    
    \subsubsection{Centering Approach}\label{naive_adjustments} 
        A key limitation of course pass rates and student GPAs in Equation \ref{course_difficulty_pass_rate} is their assumed independence from each other. For example, the GPA implicitly assumes that courses are always equally difficult (GPA weights all grades equally), while the pass rate does not consider the overall performance level of individual students. Thus, difficulty can be perceived as low when a student cohort is particularly strong. Therefore, adjustments of pass rate and GPA were introduced \citep{srivastava2024curriculum,ochoa2016simple,caulkins1996adjusting}, which center the mean course grade by the GPAs of the enrolled students and the individual student GPA by the mean course grades $\mu_c$.  
        \begin{align} \label{course_difficulty_adj_pr}
            \delta_c &= \frac{1}{|S_c|}\sum_{s \in  S} g_{s,c}-\text{gpa}_s &
            \theta_s = \frac{1}{|C_s|}\sum_{c \in  C} g_{s,c}-\mu_c
        \end{align} 
        Here, $\delta_c$ relates to the scaled difficulty of course $c\in C$ and $\theta_s$ to the performance trait of student $s\in S$. However, this adjustment may be insufficient if the adjustments ($\text{gpa}_s$ or $\mu_c$) are skewed. For example, suppose that high-achieving students systematically choose difficult courses, and low-achieving students enroll in less difficult courses. Then, the GPA as a measure of student performance would overestimate low-achieving students and underestimate high-achieving students. Thus, estimates of course difficulty based on adjustment for student GPA would underestimate the difficulty of more difficult courses and \textit{vice versa} for less difficult courses. So, it can happen that course difficulty is a concept that cannot always be calculated directly from the grades, and that needs to be be inferred as a latent factor. Because centering approaches are widely used, we use them as baseline in our evaluations.
        
        The above example leads to a further perspective: Deciding whether centering approaches are applicable requires checking their underlying theoretical assumptions. If the real-world data do not meet these assumptions, results can be misleading. Centering approaches rely on three assumptions: First, the performance of students in different courses is independent given their performance estimate $\theta_s$, i.e., $\theta_s$ captures all relevant information explaining a student's performance across different courses. For instance, this implies that the model assumes $\theta_s$ and the course selection of student $s$ to be independent. Second, course difficulty is a one-dimensional concept that neglects the idea of potential independent skills, which would require estimating multiple difficulties factors. Third, the approach assumes that courses are time-invariant and that course grades used in the GPA calculation are equally difficult. For example, the centered approach can not capture if students in one course might take less difficult courses on average than in another course, resulting in skewed difficulty estimates.
        
\subsection{Latent Variable Models} 
    To generalize the centering approach, one can assume that course difficulty is not directly observable from course grades. One must think of course difficulty as a latent concept to deal with such an assumption. This means that it must be inferred from the observable variables using statistical methodologies. Several approaches can be used to build models, depending on the type of grades captured by a dataset. If the grades are point grades on a continuous or sufficiently large metric scale (e.g., grades in $[0,100]$ or more than ten ordinal categories), they should be modeled as continuous variables. Additive grade point models are well suited for this purpose (e.g., \citealt{baucks21mitigating,caulkins1996adjusting}). Conversely, when grades are binary (e.g., pass/fail), they are modeled using logistic methods derived primarily from item response theory (IRT).  
        
    \subsubsection{Additive Model}
        The additive grade point model (AGM) \citep{caulkins1996adjusting,baucks21mitigating} follows intuitively from the centering approach of GPA and pass rates in Section \ref{naive_adjustments}. AGMs extend the idea of scaling by modeling course difficulty and student performance using statistically independent latent variables. This means that the modeled latent course difficulty is adjusted for the latent performance level of the participating students. For this purpose, it is assumed that each student's grade in a course can be modeled as the sum of the student's performance $\theta_s$ and the course's difficulty $\delta_c$:  
        \begin{align}
            g_{s,c} = \theta_s + \delta_c,
        \end{align}
        for all grades $g_{s,c}$ for student $s\in S$ and course $c\in C$. The bias terms $\theta_s$ and $\delta_c$ represent the latent trait of the student performance and course, respectively.

    \subsubsection{Item Response Theory}
        Unlike AGMs which model continuous course grades, Item Response Theory (IRT) models binary data. IRT emerged from high-stakes testing (e.g., SAT and GRE) as a response to the limitations of Classical Test Theory (CTT). CTT relies on the overall test scores of test-takers, which are analogous to student GPAs in Curriculum Analytics (CA). The test scores assume constant item properties for all items in a test. Conversely, IRT analyzes individual test items and models the probability of a correct response based on the item characteristics (e.g., difficulty) and the individual's latent performance trait. This can lead to more nuanced trait estimates because each item can behave differently.  
        
        IRT in CA models binary grades (e.g., "pass"/"fail") in courses (rather than test items) using logistic regression. Instead of modeling traits similar to AGMs bias terms, IRT models latent trait values for each student and each course that estimate the probabilities of each student passing each course. To fit the trait values, IRT maps the relation of student performance trait values and course pass rates by fitting a sigmoid function known as the item response function (IRF) for each course. The IRF maps the student's performance trait value (x-axis) to the student’s probability of passing a specific course (y-axis). Given course $c$, the position of its IRF on the $x$-axis is defined as the $x$-value where the IRF has maximum slope. This position defines the difficulty of the course, denoted as $\delta_c$. Given student performance trait $\theta_s$, and course difficulty $\delta_c$, we define the probability of passing course $c$ as:
        \begin{align}
            P(X_{s,c} = 1 | \theta_s, \delta_c) = \frac{1}{1 + e^{\theta_s-\delta_c}}.
        \end{align}
        In the literature this model is commonly referred to as Rasch model \citep{Ayala2013:Theory}.

    \subsubsection{Model Assumptions} \label{model_assumptions}
        Checking model assumptions is vital in statistical research, including education research. Unfortunately, this aspect of quantitative analyses is often neglected \citep{hoekstra2012assumptions}. Assumption checks are particularly important for robust and interpretable results. Models built on assumptions that do not hold can lead to false conclusions \citep{bergner2017measurement}. 
        AGM and IRT models employ the same three assumptions as the centering approach. 
        
        First, the \textit{unidimensionality} assumption states that latent traits of one dimension are sufficient to model the difficulty of courses and student performance. 
        To assess the suitability of this assumption, we study the number of latent dimensions required to explain variance in the student performance data and compare model fit of models that consider different dimensionality (i.e., this is possible for latent models but not for centering approaches. Centering assumes unidimensionality and is unable to handle cases where this assumption is violated). 
        
        Second, the \textit{local independence} assumption states that a student’s probability of passing a course is independent of their performance in other courses, given their latent trait. 
        \begin{align}
            P(X_{s,c}=1 | \theta_s, \delta_c, X_{s,k}) =  P(X_{s,c}=1 | \theta_s, \delta_c),
        \end{align}
        where $c,k \in C$ and $c\neq k$.
        
        Third, \textit{time-invariance} states that the fitted trait values are constant, potentially over multiple semesters and years.
        In the following sections, we discuss each assumption in detail and how to assess its applicability for both the centering approach and latent variable models. But first, we introduce some useful model extensions available for the latent models.
        
    \subsubsection{Model Extensions}
        \subsubsection*{Multidimensionality}
            In contrast to the centering approach, which is inherently unidimensional, latent variable models can be extended to model student performance trait and course difficulty via multiple dimensions.
            %We must show that the dimensionality assumption is sufficient when fitting unidimensional latent traits representing course difficulty and student performance. If not, one might ask: is the underlying latent structure multidimensional?
            IRT research shows that this can be the case, and in addition may indicate that the trait values represent multiple skills, e.g., mathematical problem-solving and text comprehension, each corresponding to separate dimensions (e.g., \citealt{hartig2009multidimensional,bacci2017evaluation}). 
            
            Again, let $C$ and $S$ be the sets of courses and students, respectively, to define the $n$-dimensional IRT model. For course, $c \in C$, the course location vector $\boldsymbol{\delta}_c \in \mathbb{R}^n$ defines the multidimensional location of its IRF over all $x$-axes. 
            However, fitting multidimensional latent traits, where each dimension of the trait affects only one specific dimension, is challenging \citep{Ayala2013:Theory}. For this reason, so-called compensatory models are used. In these models, all dimensions of the latent traits are always included in calculating the pass probability for all courses. To achieve the strongest possible separation of the dimensions of the latent traits, a discrimination vector $\boldsymbol{\alpha}_c \in \mathbb{R}^n$ is introduced, which can load the dimensions within an item. The course discrimination $\boldsymbol{\alpha}_c$ determines the slope of the IRF in each dimension. In a course $c\in C$, the probability that student $s\in S$ passes the course, i.e. $X_{s,c} = 1$, given student performance trait $\boldsymbol{\theta}_{s \in S} \in \mathbb{R}^n$, course location $\boldsymbol{\delta}_c$, and course discrimination $\boldsymbol{\alpha}_c$ is defined as
            \begin{align}\label{IRF}
                \mathbb{P}(X_{s,c} =1 \,|\, \boldsymbol{\theta}_s,\, \boldsymbol{\alpha}_c,\, \boldsymbol{\delta}_c) &\,= \frac{1}{1+ e^{-\langle \boldsymbol{\alpha}_c, \boldsymbol{\theta}_s - \boldsymbol{\delta}_c \rangle}},
            \end{align}
            where $\langle\cdot,\cdot\rangle$ denotes the Euclidean inner product. Due to the additional discrimination parameter $\boldsymbol{\alpha}_c$ for each course, this IRT model is called a two-parameter logistic model (2PL) with $n$ dimensions. We refer to it as the $2$PL-$n$Dim model~\citep{Ayala2013:Theory}. 

            We apply the same generalization from IRT research to the AGM to define the multidimensional AGM. We replace the student and course parameters with vectors. Ideally, we can cover different skills with more dimensions, analogous to multidimensional IRT. This leads to the following formulation: 
            \begin{equation}
                \begin{aligned}\label{multi_dim_agm}
                    &g_{s,c} = \langle\boldsymbol{\alpha}_c,\boldsymbol{\theta_s} + \boldsymbol{\delta_c}\rangle. %+ \epsilon_{s,c}.%, \\ 
                    %\text{s. t. } & \Cov(\theta_s^d, \theta_s^{\hat d}) = 0 \text{,  for all } d\neq \hat d, \\
                    %& \Cov(\delta_c^d, \delta_c^{\hat d}) = 0\text{,  for all } d\neq \hat d.
                \end{aligned}
            \end{equation}
            
            For both multidimensional model types, IRT and AGM, we define the single-dimensional course difficulty $\Delta_c$ of course $c$ as:
            \begin{align}\label{Difficulty}
                \Delta_c & = \frac{\langle \boldsymbol{\alpha}_c, \boldsymbol{\delta}_c \rangle}{\| \boldsymbol{\alpha}_c \|_2} \in \mathbb{R}.
            \end{align}
            The single-dimensional difficulty becomes convenient later in assessing the reliability and validity of model parameters in related experiments.
            %We will need the single-dimensional difficulty in the experiments evaluates the reliability and validity of the model parameters the corresponding experiments require a one-dimensional difficulty.
            
        \subsubsection*{Differential Course Functioning}
            IRT models and AGMs assume that the difficulty of a given course is equal for all students in the dataset. However, given fitted student performance traits and course difficulties, we may find courses for which the difficulty is not equal for students of different groups. One example might be exchange students who enter a college and struggle with the material in a particular course due to language barriers. Or we might want to study how cohorts entering a given major differ from each other in terms of their experienced difficulties. This effect is called differential functioning. IRT research tries to detect and quantify these group differences in the educational testing domain referring to it as Differential Item Functioning (DIF) (e.g., \citealt{osterlind2009differential}).
            The first application of DIF analysis in the context of university courses, referring to it as Differential Course Functioning (DCF), was done by Baucks et al.\citep{Baucks24Gaining_LAS}. The idea behind DCF is to add a covariate to the IRT model that represents students' group assignments (e.g., native vs. transfer students). If the group parameter is significantly different from zero for a particular course, the DCF effect in that course indicates disparities in the experienced course difficulty independent of the fitted student performance trait values. The same can be done analogously for the AGM.
            
\begin{figure}
                \centering
                \includegraphics[width=0.42\linewidth]{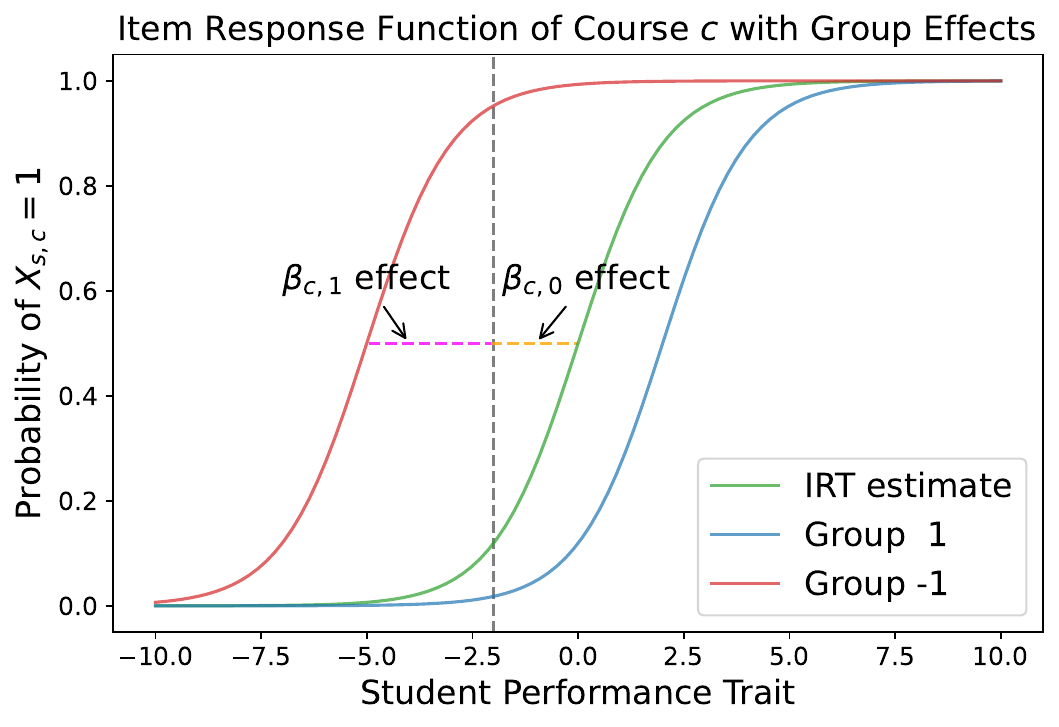}
                \scalebox{0.85}{% specifies DCF case figure
  
  \tikzset{
    % Set the overall layout of the tree
    level 1/.style = {level distance=0.cm, sibling distance=3.5cm},
    level 2/.style = {level distance=3.5cm, sibling distance=2cm},
    % Define styles for bags and leafs
    bag/.style = {text width=4cm, text centered,  inner sep=1pt},
    end/.style = {circle, minimum width=3pt,fill, inner sep=0pt}
  }
  \begin{tikzpicture}[grow=right, sloped, scale=0.625]
    \tikzset{frontier/.style={distance from root=150pt}}

    \node {}
    child {
      node[bag] (C) {$\text{PR}_{G_1} \neq \text{PR}_{G_2}$}        
      child {
        node[end, label=right:
        {\shortstack{Pot. varying trend:\\(iv) DCF $\in \mathbb{R},\, \text{PR}_\Delta \neq 0$}}] {}
        edge from parent
        node[above] {}
        node[below]  {$\theta_{G_1} \neq \theta_{G_2}$}
      }
      child {
        node[end, label=right:
        {\shortstack{Same trend:\\(iii) $\text{DCF} \neq 0,\, \text{PR}_\Delta \neq 0$}}] {}
        edge from parent
        node[above] {$\theta_{G_1} = \theta_{G_2}$}
        node[below]  {}
      }
      edge from parent 
      node[above] {}
      node[below]  {}
    }
    child {
      node[bag] (B) {$\text{PR}_{G_1} = \text{PR}_{G_2}$}        
      child {
        node[end, label=right:
        {(ii) $\text{DCF} \neq 0,\, \text{PR}_\Delta = 0$}] {}
        edge from parent
        node[above] {}
        node[below]  {$\theta_{G_1} \neq \theta_{G_2}$}
      }
      child {
        node[end, label=right:
        {(i) $\text{DCF} = 0,\, \text{PR}_\Delta = 0$}] {}
        edge from parent
        node[above] {$\theta_{G_1} = \theta_{G_2}$}
        node[below]  {}
      }
      edge from parent 
      node[above] {}
      node[below]  {}
    };    % \draw[dashed,shorten <=4pt,shorten >=4pt] (B) -- (C);
  \end{tikzpicture}}
                \caption{Adapted from 
                %([redacted], XXXX).
                \cite{Baucks24Gaining_LAS}. 
                [Left] The DCF model for a Rasch IRT framework is shown. The green sigmoid curve represents the overall response function for all students derived from the Rasch IRT model. The red and blue curves correspond to group-specific course response functions (red $\sim -1$, blue $\sim 1$), demonstrating asymmetric offsets relative to the Rasch IRT model. The parameter $\beta_{c,0}$, the intercept of the logistic regression model, quantifies the horizontal shift of the item response function (IRF) on the x-axis, which serves as an equidistant reference point for group-specific DCF IRFs. Differential difficulty between groups is captured by $\beta_{c,1}$. 
                [Right] This visualization explores potential relationships between pass rate differences ($\text{PR}_{G_1}$, $\text{PR}_{G_2}$) and DCF values for groups ($G_1$, $G_2$), considering both identical and different student performance traits ($\theta_{G_1}$, $\theta_{G_2}$). DCF allows for a deeper understanding of the specific difficulties faced by diverse student populations.}
                \label{fig:dcf_cases}
            \end{figure}
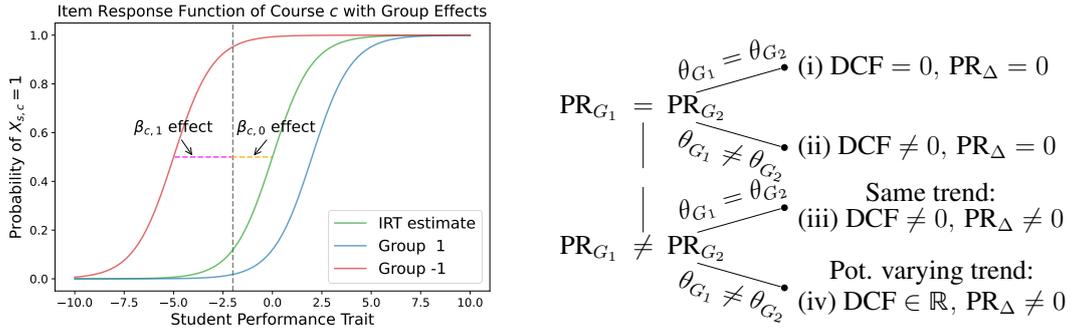

            Within the IRT framework, differential course functioning (DCF) evaluates disparities by conducting a second regression \textit{for each course} to assess potential differences between two student groups (e.g., cohort A vs. cohort B). For the fitted trait values $\boldsymbol{\theta}_s^*$ in a course $c$ we fit: 
            \begin{align}
                \text{logit}(\mathbb{P}(X_{s,c} = 1 | \, \boldsymbol{\theta}_s^*)) =  \beta_{c,0} + \beta_{c,1} g_s + \langle\boldsymbol{\beta}_{c,2}, \boldsymbol{\theta}_s^*\rangle.
            \end{align}
            Here, the logit function is the inverse of the sigmoid $\sigma (x) = 1/(1+$~$e^{-x})$. Note that the equation has no course difficulty $\delta_c$ because DCF is analyzed \textit{course by course} and is, therefore, redundant with the DCF intercept $\beta_{c,0} \in \mathbb{R}$. For AGM, we analogously fit
            \begin{align}\label{dcf_lr}
                X_{s,c} =  \beta_{c,0} + \beta_{c,1} g_s + \langle\boldsymbol{\beta}_{c,2}, \boldsymbol{\theta}_s^* \rangle,
            \end{align}
            which is essentially a linear regression.
            In both Equations, $\boldsymbol{\theta}_s^* \in \mathbb{R}^n$ is the performance trait of student $s \in S$ fitted by an initial model, IRT or AGM,  $g_s \in \{-1, 1\}$ is the DCF group encoding, $\beta_{c,0} \in \mathbb{R}$ is the DCF intercept, and $\beta_{c,1} \in \mathbb{R}$ is the DCF effect. 
            The $\boldsymbol{\beta}_{c,2} \in \mathbb{R}^n$ parameter represents the correction for the discrimination properties of the course in each dimension. It is set to $1$ in the one-dimensional case (e.g., Rasch IRT model) and varies freely in the multi-dimensional case. The detection of a DCF effect indicates that the course has systematic intergroup differences in difficulty, separate from the difficulty of the course and the fitted performance traits of the participating students.  A negative group parameter $\beta_{c,1}$ indicates that students in group $g_s = -1$ find course $c$ easier than students in group $g_s = 1$. The example, adapted from 
            %([redacted], XXXX)
            \cite{Baucks24Gaining_LAS}
            , on the left side of Figure \ref{fig:dcf_cases} visualizes that DCF example for a Rasch IRT model. The green item response function (IRF) corresponds to the model estimate, and the red ($g_s = -1$) and blue ($g_s = 1$) IRFs represent the group-specific IRFs. The purple dashed horizontal line shows the DCF effect $\beta_{c,1}$. 

            DCF provides a more nuanced approach to identifying group differences than comparing student outcomes such as pass rates (PR). For clarity and consistency with the following IRT example, we focus on PR without loss of generalizability. The right side of Figure \ref{fig:dcf_cases} presents four scenarios, adapted from 
            %([redacted], XXXX)
            \cite{Baucks24Gaining_LAS}
            , illustrating the interplay between DCF effects and pass rate differences ($\text{PR}_\Delta$) across varying mean PRs and IRT-derived student performance traits in groups $G_1$ and $G_2$. These cases, which the DCF framework can distinguish, highlight potential differences between DCF and $\text{PR}_\Delta$. Except for the null case (i), where both effects are $0$, the scenarios (ii-iv) demonstrate how DCF and $\text{PR}_\Delta$ behave differently. The cases depicted in Figure~\ref{fig:dcf_cases} are summarized as follows:
            \begin{enumerate}[label=(\roman*)]
                \item Null Effect: No evidence of disparate outcomes, as there are no differences in student performance traits ($\theta_\Delta$) or DCF effects.
                \item $\theta_\Delta$: Groups with differing performance trait levels achieve similar outcomes due to varying difficulty levels.
                \item DCF: Groups with comparable performance trait levels experience different outcomes due to differences in difficulty.
                \item $\theta_\Delta$ + DCF: Groups with differing performance trait levels experience disparate outcomes driven by both trait differences and DCF effects.
            \end{enumerate}
            In cases where the two groups differ in their underlying student trait levels, overall PRs can be confounded by the general performance gap among students. DCF mitigates this confounding by isolating course-specific difficulty effects, providing a more precise and detailed assessment of academic challenges faced by students from different backgrounds.

\subsection{Which method suits the grade scale?} \label{grade_scale}

        %Now that we have introduced the methods, the tutorial starts from a theoretical point of view. It starts with the processing of the grades.
        We first define more precisely what grade types exist. Each dataset of grades lives on a grade scale. A grade scale can exist in different forms, e.g., grades can exist as numbers or letters, or grade scales can run in opposite directions, e.g., A is best or F is best. If we have an ordinal grading scale that is not numerical (e.g., A, B,...), then we need to transform the scale into numerical form since the presented methods expect numbers. Assuming we have a numerical ordinal scale, the methods expect that grades can be measured metrically. This means, for example, that for grades $25$, $50$, and $100$, grade $100$ is twice as far away from grade $50$ as grade $50$ is from grade $25$. This scale type is called the interval scale \citep{gardner1975scales}. If this is not true, the grading scale needs to be rescaled, e.g., using a percentile transform or splitting grades into binary/dichotomous categories using the mean/median. In the following, we use the term \textit{binary} instead of dichotomous, which are synonyms originating from different research areas, i.e., machine learning and psychometrics, respectively.
        
\begin{figure}[t]
    \centering
    \includegraphics[width=0.7\linewidth]{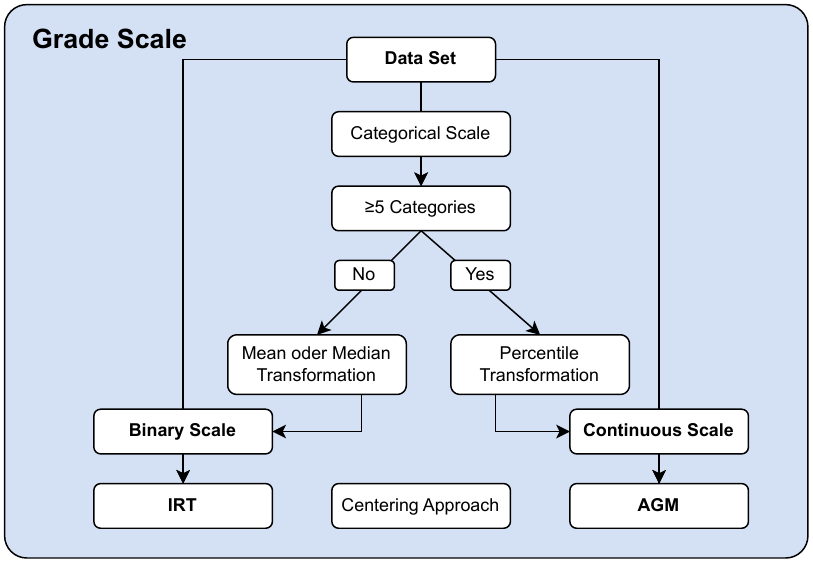}
    \caption{Decision flow for selecting the appropriate model based on grade type. This flowchart outlines the process of transforming categorical grade data and selecting between IRT models and AGMs based on whether the scale is binary or continuous. Centering procedures can be applied to both grade scales.}
    \label{fig:data}
\end{figure}

        The centering model is based on the average grades of students and courses and can be used on any interval scale. However, a distinction is critical for latent models. The AGM is based on point grade data and should, therefore, be used on interval scales that can be assumed to be continuous. If ignored, the AGM might model grades between grades nonexistent in the original scale, e.g., grades between 1 and 0 in a 1/0 (pass/fail) scale. For continuity, the scale must have a sufficient number of values. Typically, at least $5$ categories are needed to assume continuity \citep{rhemtulla2012can}. The IRT model, on the other hand, models binary data, e.g., 'pass'/'fail' grades. An overview of the model type (Centering/AGM/IRT models) selection depending on the data is shown in Figure \ref{fig:data}.

\subsection{Assumption 1: Dimensionality under Missing Data}

    \begin{figure}[t]
        \centering
        \includegraphics[width=0.81\columnwidth]{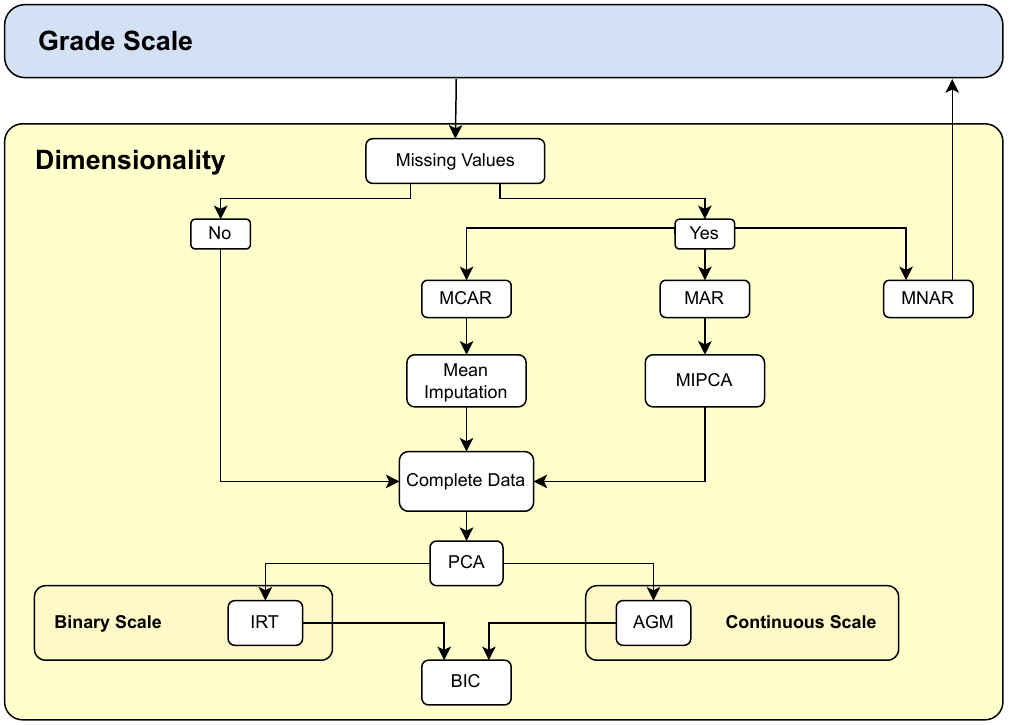}
        \caption{This flowchart illustrates the process of determining dimensionality with missing data. The approach begins by addressing missing values (if any) using MCAR or MAR assumptions and imputation techniques. Once complete data are obtained, PCA is used to determine an upper bound of the latent dimensions, followed by model selection between IRT and AGM depending on the type of grade scale, and finally, a decision on dimensionality is made using the BIC scores of the fitted IRT and AGM models.}
        \label{fig:dimensionality}
    \end{figure}
    The centering approach and the latent variable methods share the dimensionality assumption (see section \ref{model_assumptions}) that we need to test. That is the number of dimensions of the student performance trait values and course difficulty sufficiently model the data. The centering approach always assumes one dimension, while the latent variable methods can adapt to multiple dimensions if necessary. Most methods for testing dimensionality are based on complete data (i.e., no missing values) and attempt to estimate the amount of variance as a measure of information that can be explained by latent variables of different dimensions. The proportion of variance explainable by latent variables may vary depending on the research context due to dataset dependencies, e.g., by containing different noise levels or structures. Thus, finding a reasonable number of dimensions is a nuanced problem, and no rule of thumb giving thresholds for explained variance has been accepted across research domains as sufficient on its own \citep{fabrigar1999evaluating}. Within this tutorial, we tackle this problem in a two-stage process using principal component analysis (PCA) and information criteria. First, PCA evaluates how much variance orthogonal dimensions representing latent variables can capture. In the social sciences, a threshold of $50\%$ to $60\%$ is often used as a sufficient proportion of explained variance \citep{henson2006use}. The PCA results in an upper bound on how many dimensions we consider \citep{fabrigar1999evaluating}. Second, we use the Bayesian information criterion (BIC) that compares models of different dimensions to select the best tradeoff between model fit and overfitting. Figure \ref{fig:dimensionality} depicts this two-step process at the bottom. Note that missing values are addressed later in this section.  
        
    \subsubsection{Principal Component Analysis}
        Principal Component Analysis (PCA) identifies directions of greatest variance (as a measure of information) on \textit{complete} data sets. While PCA is often used for dimensionality reduction \citep{fodor2002survey}, e.g., to visualize data, it is also a valuable method for estimating the number of dimensions needed to capture most of the data's variance adequately.
        
        PCA transforms the original high-dimensional data into a new coordinate system where each axis (principal component) corresponds to a direction of maximum variance. These principal components (PC) represent the eigenvectors of the correlation matrix of the data features. In our case, the PCs represent linear combinations of the courses, capturing variance in student grades across the courses. The eigenvalues associated with these PCs indicate the variance each component captures. By analyzing the eigenvalues, we can determine the number of dimensions needed to represent the data effectively. 
        We use the correlation matrix instead of the covariance matrix here because individual courses with very high standard deviations in the grades would be over-represented proportionally by the first eigenvalue without scaling the variance. We are interested in finding individual concepts that are not correlated, so it is important to analyze courses equally.

        The largest Eigenvalues of the correlation matrix correspond to the principal components explaining most variance. Assuming we examine the Eigenvalue sizes in decreasing order. Typically, one finds an "elbow" at which the rate of decrease in eigenvalues noticeably diminishes due to the intrinsic dimensionality of the data and redundancy, e.g., due to high correlations between features. The Eigenvalue in that so-called elbow estimates an upper bound on how many PCs (or dimensions) are worth including in a later model fit. As the first step in dimensionality assessment, we apply PCA to the course grade data and estimate the number of dimensions that sufficiently explain the variance in the dataset. We ensure an efficient and informative data representation by retaining dimensions that contribute significantly to the total variance.

        Let $X \in \mathbb{R}^{|S| \times |C|}$ be a complete (i.e., no missing values) matrix with $|S|$ students and $|C|$ courses. An entry $x_{s,c}$ in $X$ represents the grade of student $s\in S$ in course $c\in C$. We define $X$ as the course response matrix. If the data is continuous, we construct the correlation matrix using the course columns in $\mathbb{R}^{|S|}$ of $X$ as variables and the Pearson correlation. 

        For binary data, constructing a correlation matrix for PCA is more complex. PCA assumes multivariate normally distributed variables.
        We can not assume variables are normally distributed if the grade scale is binary. However, we can assume that there exist variables that are continuous and normally distributed that generate the binary data, e.g., by choosing a passing threshold, we essentially generate binary data. Under this assumption, we can use binary data to estimate the correlation between the generating continuous variables. This is known as tetrachoric correlation \citep{kolenikov2004use}. We assume for each pair of courses represented by binary random variables $C_1$ and $C_2$, there exist two bivariate normal distributed variables $C_1^*$ and $C_2^*$:
        \begin{align*}
             \begin{pmatrix}
                        C_1^* \\
                        C_2^*
                        \end{pmatrix}
                        \sim N \left( 
                        \begin{pmatrix}
                        0 \\
                        0
                        \end{pmatrix},
                        \begin{pmatrix}
                        1 & \rho \\
                        \rho & 1
                        \end{pmatrix} \right)
        \end{align*}
        where $\rho$ describes the correlation between $C_1^*$ and $C_2^*$. These random variables are assumed to generate our binary variables $C_1$ and $C_2$. Then we can write: 
        \begin{align*}
            C_1 &= 
                \begin{cases}
                1 & \text{if } C_1^* > t_{C_1} \\
                0 & \text{if } C_1^* \leq t_{C_1}
                \end{cases}
            % \end{align*}
            % \begin{align*}
            &C_2 = 
            \begin{cases}
            1 & \text{if } C_2^* > t_{C_2} \\
            0 & \text{if } C_2^* \leq t_{C_2}.
            \end{cases}
        \end{align*}
        For the given cutoff thresholds $t_{C_1}$ and $t_{C_2}$, and the correlation $\rho$, the cumulative distribution function $F_{C_1^*,C_2^*}$ of the bivariate continuous random variables $(C_1^* , C_2^*)^T$ is:
        \begin{align*}
         F_{C_1^*,C_2^*}(t_{C_1}, t_{C_2}; \rho) = \frac{1}{2 \pi \sqrt{1-\rho^2}} \int_{-\infty}^{t_{C_1}}  \int_{-\infty}^{t_{C_2}} \exp\left[ -\frac{1}{2} 
         \begin{pmatrix} x \\ y \end{pmatrix}^\top \begin{pmatrix}
                        1 & \rho \\
                        \rho & 1
                        \end{pmatrix}^{-1} \begin{pmatrix} x \\ y \end{pmatrix} \right]dy\, dx 
         \end{align*}   
         Then, we can calculate the empirical probabilities of each possible case of the binary variables $C_1$ and $C_2$ and can write them as:
         \begin{equation}
         \label{tetrachoric_cdf}
            \begin{aligned}
                 \mathbb{P}(C_1=0,C_2=0) &= F_{C_1^*,C_2^*}(t_{C_1}, t_{C_2}; \rho)\\
                 \mathbb{P}(C_1=1,C_2=0) &= F_{C_1^*,C_2^*}(\infty, \infty; \rho) - F_{C_1^*,C_2^*}(t_{C_1}, t_{C_2}; \rho)\\
                 \mathbb{P}(C_1=0,C_2=1) &= F_{C_1^*,C_2^*}(t_{C_1}, \infty; \rho)-F_{C_1^*,C_2^*}(t_{C_1}, t_{C_2}; \rho)\\
                 \mathbb{P}(C_1=0,C_2=0) &= F_{C_1^*,C_2^*}(\infty, \infty; \rho)  -  F_{C_1^*,C_2^*}(t_{C_1}, \infty; \rho)-F_{C_1^*,C_2^*}(t_{C_1}, t_{C_2}; \rho).
             \end{aligned}
         \end{equation}
        For given $\rho$ we could calculate $t_{C_1}$ and $t_{C_2}$ using the inverse of the $F_{C_1^*,C_2^*}$. And for given $t_{C_1}$ and $t_{C_2}$ we could find a $\rho$ that maximizes the log-likelihood:
        \begin{align}    
            \mathcal{L}(t_{C_1},t_{C_2}; \rho)&=  \text{log} \left[\prod_{i,j \in \{0, 1\}} \mathbb{P}(C_1=i, C_2=j) n_{i,j}\right] \\
            \label{tetrachoric_logl}
            &= \sum_{i,j \in \{ 0,1 \}} n_{i,j} \log \mathbb{P}(C_1=i,C_2=j) , 
        \end{align}
        where $n_{i,j}$ is the number of occurencies. Therefore, the steps in Equation \ref{tetrachoric_cdf} and Equation \ref{tetrachoric_logl} are being done iteratively, e.g., starting with $\rho=0.5$.

        After calculating the correlations for each course pair in either way (continuous or binary), we arrive at a correlation matrix $\text{Corr}$ and can continue with PCA. Thus, we perform an eigenvalue decomposition on $\text{Corr} = V \Lambda V^T$, where $V$ is a $C \times C$ matrix of eigenvectors, and $\Lambda$ is a diagonal matrix of eigenvalues. After computing the principal components $Z = \text{Corr} V$. The grades projected onto the principal components for students are given by $Z$. The eigenvalues give the proportion of total variance explained by each principal component in $\Lambda$. Thus, we can calculate the proportion of variance explained (PVE) by the $i$-th principal component:
        \begin{align*}
            \text{PVE}_i &= \frac{\lambda_i}{\sum_{j} \lambda_j},
        \end{align*}
        where, $\sum_{j} \lambda_j$ represents the total variance across all principal components.

    \subsubsection{Missing Values}
        PCA can be applied only to complete data sets (i.e., complete course response matrices). However, missing values are common in curriculum analytics, for example, due to students dropping out or students being able to choose electives from a wide range of courses. Therefore, to apply PCA to CA datasets with missing values, we need to complete the response matrix in a process commonly called imputation. To impute missing values, we need to understand why values are missing. There are three potential types of missingness in data: Missing completely at random (MCAR), missing at random (MAR), and missing not at random (MNAR). 
        %They all share that their existence requires assumptions about their nature~\citep{schouten2021dance}. For instance, e
        Each assumes a different relationship between the missing and observed values. To impute missing values reasonably, we must assume that observed data provide sufficient information. Firstly, we state missingness as MCAR in a dataset $X \in \mathbb{R}^{|S|\times |C|}$ if the probability that values are missing is independent of the values that are observed $X_{obs}$ and the values that are missing $X_{mis}$. Let $\mathbbm{1}_{obs}$ be a matrix masking data $X$ and contain ones if values are observed and zeros if values are missing. Then the MCAR can be defined as:
        \begin{align}
             Pr(\mathbbm{1}_{obs}| X_{obs}, X_{mis}) = Pr(\mathbbm{1}_{obs})
            \end{align}
        Imputation for MCAR data does not usually bias the results, as the probability of being missing is the same for all data. An example is system errors, such as grades randomly not being entered into the system. Secondly, missing values are MAR if the missingness is dependent on the observed values $X_{obs}$ but independent of the missing values $X_{mis}$.
        \begin{align}\label{mar}
             Pr(\mathbbm{1}_{obs}| X_{obs}, X_{mis}) = Pr(\mathbbm{1}_{obs}| X_{obs})
        \end{align}
        Imputation can be safely performed under the MAR assumption if the imputation model accurately accounts for the variables driving the missingness. An example is when the missingness is related to dropout, which could be driven or explained by low performance. Lastly, we call missing data MNAR if Equation \ref{mar} is not fulfilled. This means the existence of missing values, which is not random, but in \textit{addition}, systematically related to the missing values themselves $X_{mis}$. For example, grades could be MNAR if they are manipulated a posteriori to be missing because they were too low on average. 
        
        Choosing the proper imputation methods for different types of missingness is essential \citep{howard2015using}. Failure to do so can potentially introduce bias into the results \citep{schouten2021dance} and thus lead to false actions by stakeholders relying on the biased insights. 
        
    \subsubsection*{Identification of Missing Value Type}
       In the CA context, i.e., grades in university courses, it makes sense to consider beforehand how missing values can occur and, based on that, which types come into question. We believe that MAR is usually present because, e.g., students drop out based on low performance and, therefore, have missing grades from courses taken later in their studies. Since GPA and course grades are known to be significant predictors of dropping out (e.g., \citealt{gershenfeld2016role}), we would expect MAR to be present here. However, these statements should never be taken as absolute, as it is impossible to reproduce all variation in the data \citep{boeve2019natural}. Students might also drop out independently of their grades but because of other aspects not represented in the data, indicating MCAR. We will provide methods for testing and imputation of MCAR and MAR values in the following and the Appendix \ref{app:little}.
       
    \subsubsection*{Little's test for MCAR}
        Little's test \citep{little1988test} assesses whether the missing values in the data are MCAR by examining whether the pattern of missingness has detectable structure. The test works by assuming that, under MCAR, the means of the observed values should be similar across different patterns of missingness. For each pattern, we compare the expected mean (based on the assumption of MCAR) with the actual mean observed in the data. If these means are significantly different, Little's test suggests that the data are likely not MCAR, i.e., the missingness may depend on the data values themselves. A low p-value ($<0.05$) indicates that MCAR is unlikely to be a valid assumption. A detailed theoretical derivation is provided in Appendix \ref{app:little}.

    \subsubsection*{Predicting Missingness for MAR and MNAR} \label{pred_miss}

        If Little's test indicates that the missing data is unlikely MCAR, we next want to test whether the data is MAR. According to the definition of MAR in Equation \ref{mar}, we need to show that we can explain the probability of missingness to a significant degree using the non-missing grades. To do this, for each course in the data, we fit a logistic regression model that predicts whether a grade in that course will be missing given the students' GPAs, grade standard deviations, grade minimum, and grade maximum of all other courses. These features can capture the basic properties of the student's grade distribution, such as position and outliers. If the observed grades explain the missingness of the target course, the fitted parameters are significantly different from zero, supporting MAR. In addition, McFadden's pseudo $R^{2}$ \citep{veall1996pseudo} value is reported to assess a relative measure of the variance the models explain. Unlike the $R^2$ value used in linear regression, pseudo $R^2$ is not a proportional measure of explained variance. It is not expressed as a percentage like $R^2$. The McFadden pseudo $R^2$ is generally lower than a continuous $R^2$ and increases monotonically with added variables. McFadden describes a value of $0.2$-$0.4$ as indicating an excellent model fit \citep{mcfadden1974conditional}. However, values less than $0.2$ are common and often still indicate a meaningful model \citep{ugba2022modification}. We, therefore, flag a model fit with a pseudo $R^2<0.1$ as being at risk of not providing enough evidence for MAR, thus indicating the need to interpret these values with care in the following analyses.  

        If we cannot find a statistically significant relationship between the observed and missing values, we do not have enough evidence to rule out MNAR. If MNAR is present, i.e., the missingness depends on the missing values, then standard imputation methods, such as multiple imputation, are not readily applicable. In this situation, the best way to proceed is to collect the missing data or other new data that can explain the missingness. 

    \subsubsection{Imputation of Missing Values} \label{mipca_imputation}
        Assume the previous analyses indicated either MCAR or MAR for the missing values. Then, we want to impute the course response matrix to move forward with PCA. The imputation depends on the type of missing values. In the case of MCAR, we can use simple mean or median imputation, which is defined as $x_{i,j} = \hat\mu_{obs}$, and $x_{i,j} = \hat m_{obs}$, where $\hat\mu_{obs}$, $\hat m_{obs}$ are the empirical mean and median, respectively, on the observed data. 
        
        In the case of MAR, we use an iterative imputation method called multiple imputation PCA (MIPCA) \citep{josse2016missmda} for continuous data and the tetrachoric correlation adjusted PCA in MIPCA for binary data. MIPCA uses principal component analysis to learn low-dimensional representations of courses on the available data and uses them to impute the missing values multiple times. MIPCA begins by imputing missing values under the Missing Completely at Random (MCAR) assumption using mean imputation. PCA is then applied to the imputed data set to estimate principal components. These PCA estimates are then used to generate updated imputations for the missing values. This process is iterated until convergence is reached, ensuring consistent estimates of both the principal components and the missing data.
        %Using PCA, MIPCA effectively handles multicollinearity in the data by transforming correlated variables into uncorrelated principal components. This is particularly useful in our application because we expect and want to model lower-dimensional representations of students and courses.         

    \subsubsection{Reliability of Explained Variance under Imputation}
        To ensure PCA remains reliable with missing data, we test how varying rates of missing values (assumed to be MAR) affect the explained variance in dimensionality assessment. Since imputation can distort PCA’s variance explanation, we simulate complete datasets and then "mask" values under MAR conditions, as detailed in the Appendix \ref{app:reli_amputation}.
        
        We simulate realistic dropout patterns, where missingness likelihood depends on student performance and course difficulty. We set masking rates for each simulation depending on student performance and course difficulty, generating various global masking rates for each scenario. After masking, we apply both mean imputation (for MCAR) and MIPCA (for MAR) to restore missing values. We then compare the variance explained by PCA in both the imputed and original datasets. If imputation is effective, the variance explained should remain stable. Our results in Appendix \ref{app:reli_amputation} show that MIPCA closely preserves the true explained variance under MAR, while mean imputation underestimates it—emphasizing the importance of MAR-specific imputation methods for reliable PCA.
        
    \subsubsection{Bayesian Information Criterion}

        The Bayesian Information Criterion quantifies the trade-off between model fit (log-likelihood) and potential overfitting (number of model parameters) and is a form of in-sample validation which is desirable in many CA applications where sample sizes are limited.
        
        After selecting an appropriate latent variable model according to the corresponding grade scale of the data, an upper bound on the number of latent dimensions is determined using PCA. When multiple dimensions are possible (e.g., PCA indicates two latent dimensions), we need a criterion to compare the potential models with different dimensionalities relative to each other. For this we employ the Bayesian Information Criterion (BIC) \citep{Ayala2013:Theory}. The BIC balances model fit, as measured by the log-likelihood, against the risk of overfitting by penalizing the number of model parameters. It serves as a type of in-sample validation. For two models to be comparable, BIC requires that the model of one dimension be nested within its higher-dimensional version, like the polynomial of degree two is nested in the polynomial of degree three.
        
        The BIC requires that the parameter spaces of a model of one dimension be nested within the parameter space of its higher dimensional version. For IRT and AGM models, this is always true and we can compare IRT and AGM models of varying dimensions in their respective model classes against each other (note, we can \textit{not} compare AGM vs. IRT models as there parameter spaces are not nested.). To define the BIC, assume we have fitted a model $M$ such that model parameters $\hat \theta$ maximize the model's likelihood $\hat L = p(X_{obs} | \hat \theta, M)$.
        Then, we define the BIC:
        \begin{align}
            BIC  =  k ln(S) - 2 ln(\mathcal{\hat L}),
        \end{align}
        where $k$ is the number of parameters of model $M$, and $S$ is the number of data points (e.g., number of students).
         This will help us decide which model, and therefore which data dimensionality, is appropriate for further analyses. We need the likelihoods $\mathcal{\hat L}$ in their analytical form to calculate the BIC scores of the models. These are derived in Appendix \ref{app:likelihoods}.

\subsection{Assumption 2: Local Independence}
    The second central assumption shared by all three modeling approaches is local independence (LI). Local independence states that students' performance in all courses is independent, given their performance trait values:
    \begin{align}
        P(X_{s,1}, X_{s,2}, ..., X_{s,C} | \theta_s ) = \prod_{i=1}^{C} P(X_{s,i} | \theta_s)
    \end{align}
    The assessment of LI is inherently complex because it requires understanding both the observable patterns in the data and the underlying theoretical concepts the courses are supposed to measure. The most common criterion, Yen's Q3~\citep{Yen1993:Scaling}, leverages residual correlations to give a necessary but non-sufficient criterion for local independence. Thus, Yen's Q3 can identify course pairs at risk of violating LI but can not guarantee course pairs to be LI. Residual correlation is measured by the Pearson correlations between the residuals of the courses, i.e., the difference between the grade and the model estimate. If the grades are binary (i.e., 'pass/fail'), we use the difference between the grade and the modeled \textit{pass probability} to achieve continuous residuals. The residuals should be normally distributed around $0$ if the LI assumption holds. If LI is violated, i.e., the fitted model parameters do not exclusively explain the parameters of the courses, systematic information remains in the residuals, which the Pearson correlation can measure. Mathematically, Yen's Q3 is defined as the correlation between the residuals of two courses across all students. Specifically, if \(r_{ij}\) is the residual for course \(i\) for examinee \(j\), and \(n\) is the total number of examinees, then Yen's Q3 between course \(i\) and course \(k\) is calculated as follows:
    \begin{align*}
        Q3_{ik} = \frac{\sum_{j=1}^{n} (r_{ij} - \bar{r}_i)(r_{kj} - \bar{r}_k)}{\sqrt{\sum_{j=1}^{n} (r_{ij} - \bar{r}_i)^2} \sqrt{\sum_{j=1}^{n} (r_{kj} - \bar{r}_k)^2}}
    \end{align*}
    where \(\bar{r}_i\) and \(\bar{r}_k\) are the mean residuals for courses \(i\) and \(k\), respectively. High values of Q3 suggest a significant residual correlation, thereby indicating violations of the local independence assumption, while values close to zero suggest that the assumption may hold.

    In real-world data, correlations can be expected to occur to a small extent because models cannot account for all the natural variation in the data \citep{boeve2019natural}. Therefore, guidelines for critical correlation values exist in the literature \citep{christensen2017critical}. These are set relative to the Q3 average of all course pairs. Following guidelines, we consider the assumption at risk when the Q3 value of a pair of courses differs by more than $0.2$ from the average Q3 value across all pairs of courses \citep{christensen2017critical,Ayala2013:Theory}. When this happens and the residual correlation is positive, the corresponding course pair needs to be combined into one course by taking the rounded mean grade. This is \textit{not} done automatically by the software package. Then, the Q3 computation is repeated until no more pairs are above the threshold. Alternatively, if not combined, the course estimates must be interpreted cautiously in downstream analyses. 
    
    The LI assumption is closely related to the dimensionality assumption \citep{chou2010checking}. Explaining a large amount of the variance in the data using a model of a given dimension leads to most course pairs being locally independent. However, the LI and dimensionality assumptions are not the same \citep{Ayala2013:Theory}. A dataset might correspond to a one-dimensional student performance trait but contain more complex nonlinear dependencies between single pairs of courses that the trait can not capture.
    %One might have hierarchical courses built on top of each other. Then, these courses would violate the LI assumption but not the dimensionality assumption. 
    Finally, if most course pairs violate the LI assumption, this may also indicate that the model does not represent the underlying latent structure measured by PCA. This can be tested by applying PCA to the model residuals of the imputed data set and comparing the resulting variances with those resulting from PCA applied to the imputed data (e.g., \citealt{chou2010checking}).

   %Why Yen's Q3 is not sufficient: Pairwise comparison cannot capture complex interdependencies, but also pairwise dependencies can be crazy complex such that they can not be captured by traits of an IRT model. One way to handle that is assessing the dimensionality of the data since it gives us reason to believe that the underlying latent dimensions are correctly specified increasing the confidence in the IRT parameter estimates. Or if independence is at risk one can group courses and see them as one single grade. In the end, we are interested in building confidence in our approximating assumption by leveraging different statistical methods.

\subsection{Assumption 3: Time-invariance of Course and Student Parameters}\label{assumption_time}
    We assume that the student performance traits and course difficulty fitted by the models are constant over time. This is not straightforward since course difficulty can change over time \citep{baucks24lak} and one could assume a learning rate for student ability values \citep{koedinger2023astonishing}. The constant student performance and course trait parameters can not represent such change.
    \begin{figure}[t]
        \centering
        \includegraphics[width=0.32\linewidth]{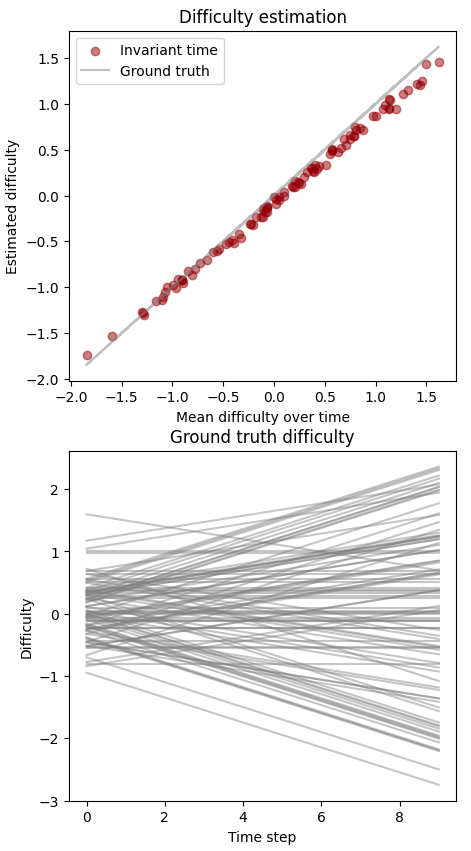}
        \includegraphics[width=0.32\linewidth]{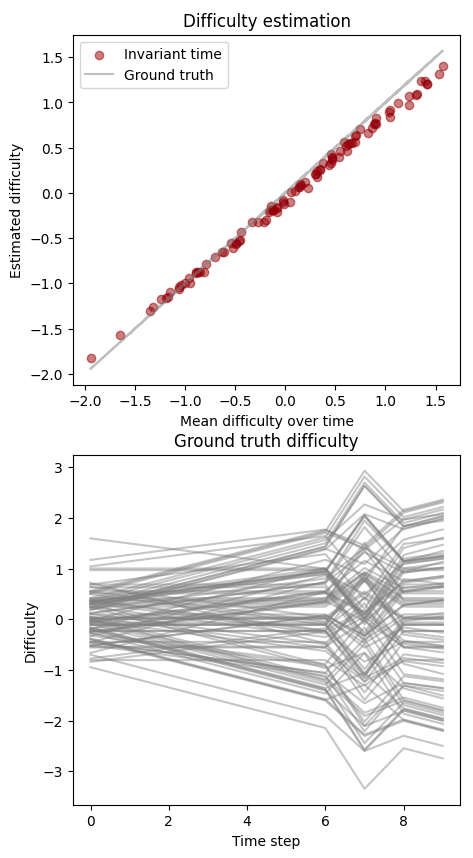}
        \includegraphics[width=0.315\linewidth]{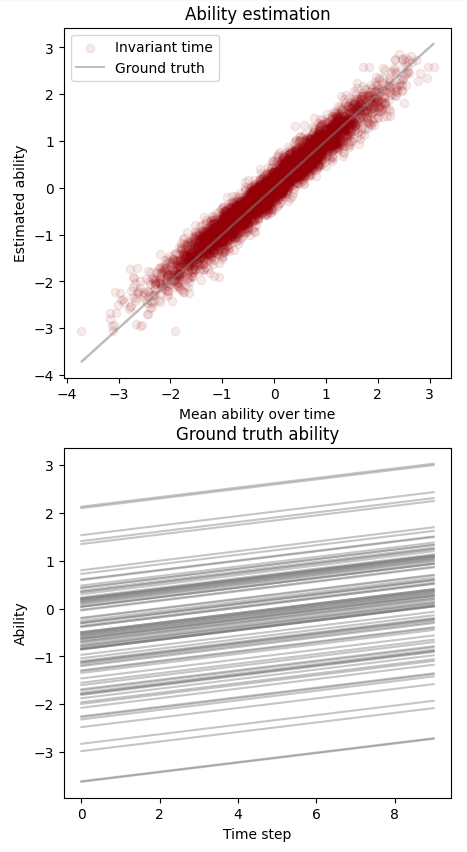}        
        \caption{Simulation study of the effects of violating the time-invariance assumption. When time drift is present, the mean is well-fitted for both courses and students. On the \textbf{left}, the bottom plot shows simulated difficulty drifts for each \textit{course} at a constant rate, where each gray line represents a course. The top scatterplot compares the mean difficulty of each course ($x$-axis) with the IRT-fitted difficulties ($y$-axis) and shows a high correlation of $0.999$ ($p < 0.001$). In the \textbf{middle}, the bottom plot shows simulated difficulty drifts with an additional shock at the 7th time step, randomly changing the course difficulty. The top scatterplot, again comparing mean difficulty to IRT-fitted difficulty, maintains a high correlation of $0.999$ ($p < 0.001$). On the \textbf{right}, the simulation models a drift in \textit{students}' latent performance traits over time at a constant rate. The top scatterplot compares mean student performance trait values ($x$-axis) with IRT-fitted student performance traits ($y$-axis) and shows a high correlation of $0.974$ ($p < 0.001$).}
        \label{fig:invariance_assumption}
    \end{figure}

    We simulate three data sets to examine the robustness of the model fit to changing traits over time. Two datasets simulate changing course traits, and one simulates changing student performance traits. We simulate grades using a ground truth IRT model and normally distributed latent traits. Then, the latent traits are modified as follows, resulting in the three data sets (c.f., Figure \ref{fig:invariance_assumption}): (i) course difficulty changes constantly, (ii) course difficulty constantly changes \textit{and} with an outlier difficulty in a single time step, and (iii) student performance trait changes constantly with the same rate for each student according to \cite{koedinger2023astonishing}. Here, we limit the results to the IRT model, as these are expected to generalize to AGMs. Figure \ref{fig:invariance_assumption} shows the changing traits over time and the corresponding IRT models' estimates that are constant over time. 
    %In the simulation, it is apparent that the best model fit .... is obtained when the mean over time of the simulated ground truth traits is fitted.
    In all three simulation results (top), it is evident that the latent traits of the optimized model ($y$-axis) are similar to the mean trait over time of the simulated ground truth traits ($x$-axis).

\begin{figure}[t]
            \centering
            \includegraphics[width = \textwidth]{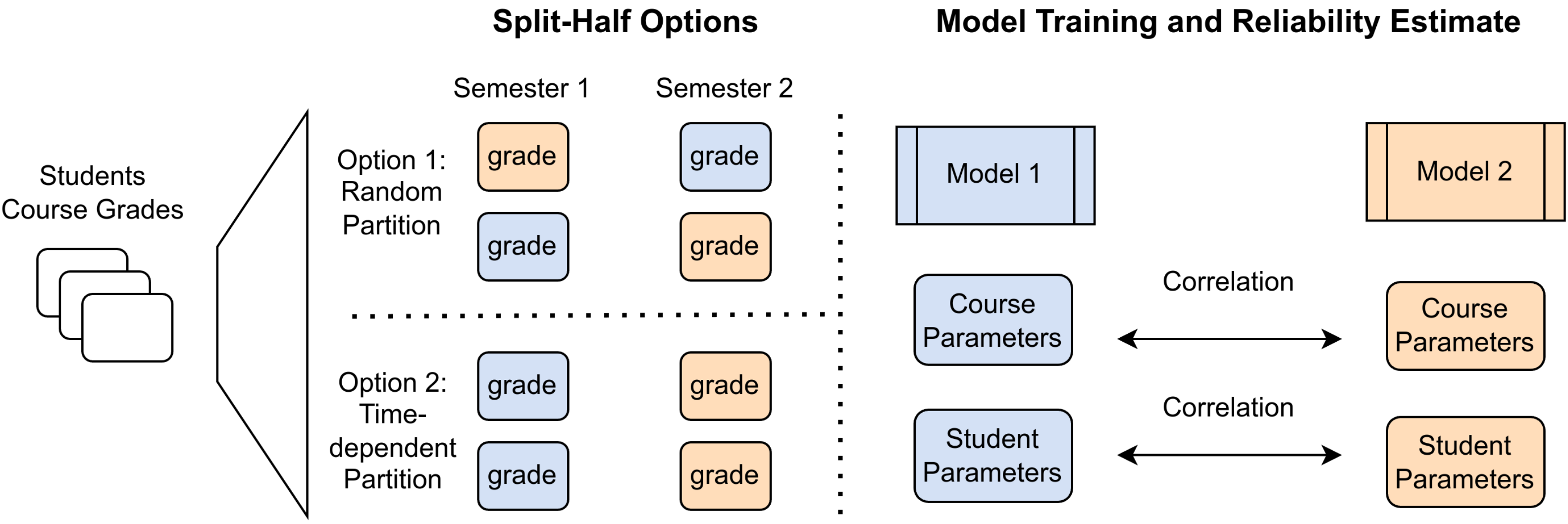}
            \caption{Split-half testing procedures for reliability and parameter time-invariance assessments. Student grades are partitioned in one of two ways: (1) randomly or (2) time-dependent. In the time-dependent method, courses are sorted by semester, with the first half assigned to one dataset and the second half to another. Two separate models are trained on each partition, and the resulting student performance and course difficulty trait values are compared using Pearson correlation.}
            \label{fig:split_half}
    \end{figure}

    When time invariance is violated, we have two options: we use the model as is and are satisfied with only being able to model the mean, or we model a course in each semester as a separate course, called course offering. But the latter is only possible when we have enough students in each course ($>75$ students per course offering) \citep{baucks24lak}.
    For students, the number of courses they attended over time is typically larger than the number a course was offered over time. We do a split-half reliability test to test for a drift in student performance. If we get two different means, that would indicate that there actually is a drift \citep{baucks24lak}. Here, split-half testing sorts the student's grades by time and then partitions each student's grades into the first and second half. This results in two distinct datasets, each containing each student (see Figure \ref{fig:split_half}). After fitting two distinct models, one on each dataset, we can compare the fitted student performance trait values. If they are very similar, we can assume stable parameters. If they are not similar, we can limit our interpretation of the trait values to statements about the mean.
    
    \subsection{Reliability and Validity} 
        Once we have chosen a model and checked its assumptions, we need to verify that the fitted course and student parameters are valid and reliable. Validity means that the parameters capture the concepts we want to model, i.e., course difficulty and student performance. Reliability means that the parameters are robust to re-fitting the model on resampled data. Both concepts are essential for generating trustworthy CA insights.
        
        \subsubsection{Concurrent Validity}
            For validity, we test the concurrency of model parameters by comparing the model's latent trait parameters to variables that attempt to measure the same concept (e.g., course GPA, which measures course difficulty). To examine concurrent validity, we assess the relationship between course difficulty and course average grade, as well as student performance parameters and GPA, using Pearson correlation. High correlation values point to a strong relationship, indicating valid parameters.
        \subsubsection{Internal Consistency Reliability}
            We assess reliability by checking the consistency of the model parameters using an internal consistency approach with split-half testing. Unlike the time-based partitioning in section \ref{assumption_time}, this test randomly splits the dataset into two disjoint sets (see Option $1$ in Figure \ref{fig:invariance_assumption}). We then assess whether the model produces comparable results on the two sets.
            For internal consistency reliability, we fit independent models to each set. Consistency is quantified using the Pearson correlation between the model parameters from each subset. First, the sets of course parameters are compared, and second, the sets of student parameters are compared. We expect high correlation values if the model fit is reliable.

\section{Case Study} \label{sec:applications}
\begin{figure}
            \centering
            \includegraphics[width=0.81\columnwidth]{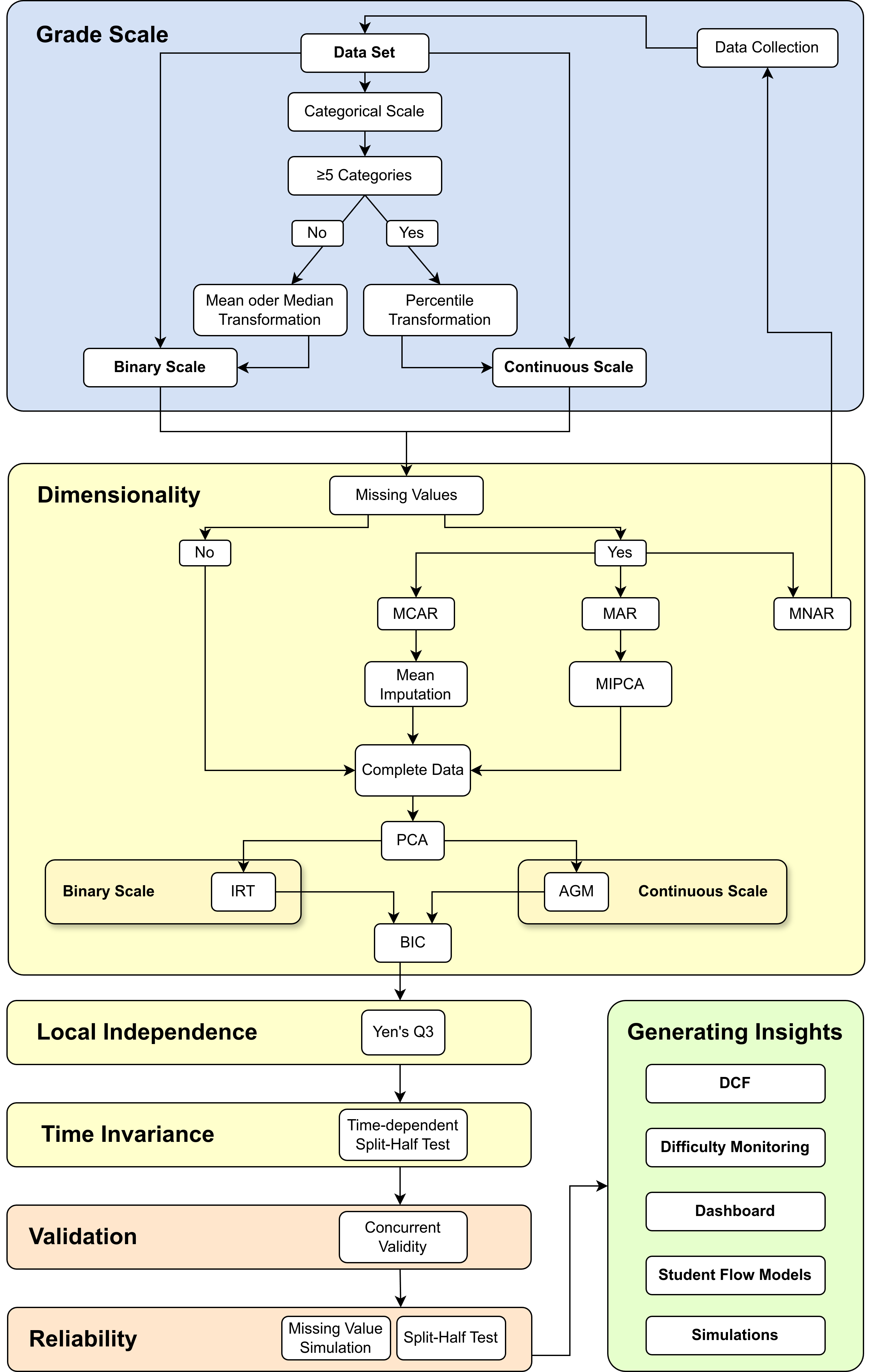}
            \caption{Flowchart outlining the process for using IRT and AGMs in course difficulty estimation. The process is divided into several key steps: In the blue box, grade scale transformation is situated. The yellow boxes represent the model assumptions: dimensionality analysis (including missing value imputation), local independence checks (using Yen's Q3), and time invariance tests (using time-dependent split-half test). The orange boxes represent the assessment of measurement properties: validation (using concurrent validity) and reliability assessment (using split-half test and missing value simulation). Finally, the green box generates insights (with applications such as DCF, difficulty monitoring, dashboards, student flow models, and simulations).}
            \label{fig:method}
        \end{figure}

        In the following, we apply our methodology to different data sets: two simulated and two real-world datasets. We summarized the entire methodology in Figure \ref{fig:method} as a flowchart. The flowchart allows users to decide which method is most suitable for its application, how to test its assumptions, how to assess reliability and validity, and finally, decide on insights that can be generated. The 'course difficulty estimation' (CDE) package can automatically select a model and test its assumptions if the tutorial in Section \ref{sec:hands_on} is followed.  We will use the CDE package to address research questions related to the influence of external events, group differences, and degree fairness.

    \subsection{Data Sets} \label{sec::data_sets}
        \subsubsection{Real-World Data}
            This study uses two different real-world datasets previously introduced in our IRT study 
            \citep{baucks24lak}
            %([redacted], XXXX)
            , both of which capture multiple years of academic performance at a German university in the 3-year Computer Science (CompSci) and Mechanical Engineering (MechEng) undergraduate programs.
    
            The CompSci dataset includes exam results of 1,098 students in 19 compulsory courses from 2013 to 2021. Each course is evaluated on a grade scale of 0 to 100, with a passing threshold of 50. Each grade was determined by a single end-of-semester exam. All identifying information was removed to preserve privacy, and a uniform stochastic noise of ±5 points was applied to each grade. For data consistency, only first-time exam attempts were included, excluding retakes, and students with fewer than five non-zero exam grades were dropped, leaving a final sample of 664 students.

            The MechEng dataset consists of exam results from 3,059 students across 18 compulsory courses from 2012 to 2021. The original grading system ranged from 5.0 (fail) to 1.0 (pass), with unequal intervals between grades. Again, each course grade was determined by a single end-of-semester exam. Following the methodology in section \ref{grade_scale}, we standardize the data by putting it on a ratio scale. Since the grades contain more than $5$ ordinal categories, we applied a percentile transformation to convert the grades to a ratio scale from 0 to 100, where higher numbers indicate better performance. As with the CompSci data, anonymization was applied, and only first-attempt exams were retained. After processing, the final sample consisted of 1,651 students.
    
            Finally, we duplicated and transformed each dataset by converting the point grade data to binary data using a 'pass'/'fail' conversion. Thus, we have each of the two real world datasets twice: two distinct point grade datasets on a [0,100] continuous ratio scale and two distinct binary datasets on a 'pass'/'fail' scale. Following our methodology (see Figure \ref{fig:method}), we would have used \textit{only} the continuous grade data to preserve as much information as possible. However, we transformed the data to binary format in order to demonstrate the IRT model, too. Thus, when we apply the IRT model to the dataset, we use the binary data; otherwise, we use the continuous scaled data. 

        \subsubsection{Simulated Data for Baseline and Upper Bounds}
            We have introduced many ideas and statistics in the Assumption Testing, Reliability, and Validation sections, which are subject to so-called critical values. For example, for local independence, a Q3 value difference of $0.2$ from the average Q3 value indicates that the assumption is at risk. To also get an idea of an upper bound (how good can our results be under the best conditions?), we simulate the selection process and the validity/reliability tests using simulated data in two ways: (i) we use a ground truth one-dimensional IRT model, and (ii) a ground truth two-dimensional IRT model based on normally distributed student ($|S|=2000$) and course traits ($|C|=10$) to generate 'pass'/'fail', and point grade data. We scale IRT's simulated pass probabilities to a [0,100] scale for point grade data. Repeating the data generation $10$ times, we simulate $10$ datasets for each dimensionality and grade scale. We report all results as the \textit{mean} over the $10$ datasets in each setting.
            %Thus, we use the binary data whenever we apply the IRT model to the dataset; otherwise, we use the continuous scaled data. 

    \subsection{Assumption Checking}
        
        \subsubsection{Assumption 1: Dimensionality}
            Following the methodology in Figure \ref{fig:method}, we continue by assessing the first assumption, dimensionality. Since the dimensionality assessment requires complete datasets, we ask if missing values are apparent and which type of missingness is apparent. The simulated datasets are simulated without missing values. However, both real-world datasets show missing values in each course at rates less than 44\% in CompSci and less than 29\% in ME. For dimensionality testing, we use the continuous versions of the datasets.

            \paragraph{Assessing Type of Missingness}
            We apply Little's test to test for missing completely at random (MCAR). For both datasets, CompSci and MechEng, Little's test results in $p$-values larger than $0.05$, meaning there is insufficient evidence to state that the missing values are MCAR. One might falsely conclude that the missing values are not MCAR, which is not what the test states. 
            Instead, we try to find statistical evidence for MAR by following our methodology for characterizing missingness. We use the students' grade distribution characteristics (i.e., GPA, standard deviation, minimum, and maximum) as predictors of missingness in logistic regression for each course. If the distribution characteristics contribute significantly to the prediction and the pseudo $R^2$ of the logistic regression model is larger than $0.1$, we assume MAR. In Table \ref{tab:mar}, we report the logistic regression results using $p$-values indicating if the predictor coefficients significantly differ from zero and pseudo $R^2$.

            For CompSci, the regression coefficient for GPA is often not significantly different from zero, where coefficients of standard deviation, minimum, and maximum grades differ significantly from zero for most courses. Thus, the variance in students' grades seems to be a better predictor of missingness than the position of the grade distribution. However, the results show that significant predictors for missingness exist in the non-missing grades, indicating potential MAR. To be confident about MAR, we show that the given predictors explain sufficient information using the pseudo $R^2$. Again, pseudo $R^2$ values for each course are reported, showing that in most courses,  pseudo $R^2$ values are above the $0.1$ threshold given in Section \ref{pred_miss}, indicating sufficient model fit. The courses 'Statistics', 'Economics', 'SoftEng', and 'WebEng' have $p$-values larger than $>0.05$ for every predictor and a pseudo $R^2<0.1$ indicating insufficient fit under the given model. This does not always mean the missingness is MNAR, but it could mean that we have not found variables that explain enough variance in the missingness. However, we must remember that the models fitted in downstream analyses will likely not capture all relevant aspects of the course difficulty and student performance, especially concerning dimensionality. Thus, these courses need to be interpreted with caution.

            For MechEng, the pseudo $R^2$, generally, seems to be lower than for the CompSci courses and is less often above $0.2$, indicating that the variables explain less missingness. This is likely due to the grade scale of MechEng. In the original scale, only the grade of $5.0$ indicates a failing, whereas all other grades indicate a passing grade. The preprocessing condition of at least $5$ grades $>0$ for MechEng is equivalent to CompSci demanding $5$ grades $>50$ for each student, thus filtering out more students with lower grades. Similar to CompSci 4/18 courses show small pseudo $R^2$ values $<0.1$. Again these courses, 'Chemistry', 'Mathematics II', 'Mechanics II', and 'IndustrialMgmt' need to be interpreted cautiously.
             \begin{table}
    \centering
      \caption{Predicting missingness with logistic regression. Each student's grade distribution features (i.e., mean, deviation, minimum, and maximum) are significant predictors of missingness. Sufficient pseudo-$R^2$ values ($>0.1$) suggest that the MAR assumption is reasonable, which holds for most courses in both majors (CompSci on the left and MechEng on the right).}

            \begin{minipage}[t]{0.49\textwidth}
                \resizebox{\textwidth}{!}{
                    \begin{tabular}{l|ccccc}
                        \toprule

                        \textbf{CompSci} Courses  &  GPA & STD & MIN & MAX &  pseudo r2  \\
                                    \midrule
                        CompNets  &  0.00 &  0.00 &  0.00 &  0.00 &  0.55  \\
                        Mathematics I  &   \textcolor{purple}{0.29}   &  0.00 &  0.00 &  0.00 &  0.28  \\
                        Mathematics II  &  0.00 &  0.00 &  0.00 &  0.00 &  0.53  \\
                        CompSci I  &   \textcolor{purple}{0.25}   &  0.00 &  0.00 &  0.00 &  0.23  \\
                        CompSci II  &  0.00 &  0.00 &  0.00 &  0.00 &  0.52  \\
                        ObjModeling  &  0.00 &  0.00 &  0.00 &  0.00 &  0.6  \\
                        Programming  &  0.00 &  0.00 &  0.00 &  0.00 &  0.23  \\
                        Statistics  &   \textcolor{purple}{0.62}   &  0.01  &   \textcolor{purple}{0.07}   &   \textcolor{purple}{0.34}   &   \textcolor{purple}{0.04}   \\
                        Privacy  &  0.00 &  0.00 &  0.00 &  0.00 &  0.20  \\
                        Economics  &   \textcolor{purple}{0.21}   &   \textcolor{purple}{0.05}   &   \textcolor{purple}{0.82}   &   \textcolor{purple}{0.05}   &   \textcolor{purple}{0.07}   \\
                        Databases  &   \textcolor{purple}{0.77}   &  0.00 &  0.00 &  0.00 &  0.27  \\
                        Data Structures  &   \textcolor{purple}{0.94}   &  0.00 &  0.00 &  0.00 &  0.27  \\
                        DiscMath  &  0.01  &  0.01  &   \textcolor{purple}{0.70}   &   \textcolor{purple}{0.09}   &   \textcolor{purple}{0.07}   \\
                        Management  &  0.00 &  0.00 &  0.00 &  0.00 &  0.58  \\
                        CompArch  &  0.00 &  0.00 &  0.00 &  0.00 &  0.69  \\
                        SoftEng  &   \textcolor{purple}{0.74}   &   \textcolor{purple}{0.19}   &   \textcolor{purple}{0.16}   &   \textcolor{purple}{0.68}   &   \textcolor{purple}{0.03}   \\
                        CompSci III  &  0.00 &  0.00 &  0.00 &  0.00 &  0.56  \\
                        OpSys  &  0.00 &  0.00 &  0.00 &  0.00 &  0.59  \\
                        WebEng  &   \textcolor{purple}{0.46}   &   \textcolor{purple}{0.82}   &   \textcolor{purple}{0.91}   &   \textcolor{purple}{0.61}   &   \textcolor{purple}{0.01}   \\
                        \bottomrule
                \end{tabular}
                }
                
            \end{minipage}
                \hspace{0.01cm}
            \begin{minipage}[t]{0.49\textwidth}
                \resizebox{\textwidth}{!}{
                    \begin{tabular}{l|ccccc}
                        \toprule
                        \textbf{MechEng} Courses  &  GPA & STD & MIN & MAX &  pseudo r2  \\
                        \midrule
                        % Business  &   \textcolor{purple}{0.193}   &  0.00 &  0.00 &  0.00 &   0.192\\
                        % Chemistry  &   \textcolor{purple}{0.987}   &  0.00 &  0.00 &  0.00 &   0.136\\
                        % ETech  &  0.008  &  0.00 &  0.00 &  0.00 &  0.259  \\
                        % ControlTech &   \textcolor{purple}{0.501}   &  0.00 &  0.00 &  0.00 &   0.176\\
                        % FlowMech  &   \textcolor{purple}{0.118}   &  0.00 &  0.00 &  0.00 &  0.205  \\
                        % IndMngmt  &   \textcolor{purple}{0.183}   &  0.00 &  0.00 &  0.00 &   0.127\\
                        % Construction I  &   \textcolor{purple}{0.064}   &  0.00 &  0.00 &  0.00 &   0.191\\
                        % Construction II  &  0.011  &  0.00 &  0.00 &  0.00 &   0.176\\
                        % Mathematics I  &   \textcolor{purple}{0.495}   &  0.00 &  0.00 &  0.00 &   0.104\\
                        % Mathematics II  &   \textcolor{purple}{0.951}   &  0.00 &  0.00 &  0.00 &   0.136\\
                        % Mathematics III  &   \textcolor{purple}{0.406}   &  0.00 &  0.00 &  0.00 &  0.222  \\
                        % Mechanics I  &   \textcolor{purple}{0.058}   &  0.00 &  0.00 &  0.00 &   \textcolor{orange}{0.084}\\
                        % Mechanics II  &   \textcolor{purple}{0.766}   &  0.00 &  0.00 &  0.00 &   0.179\\
                        % NumMath  &   \textcolor{purple}{0.179}   &  0.00 &  0.00 &  0.00 &  0.285  \\
                        % Physics  &   \textcolor{purple}{0.251}   &  0.00 &  0.00 &  0.00 &   \textcolor{orange}{0.085}\\
                        % Thermo  &   \textcolor{purple}{0.667}   &  0.00 &  0.00 &  0.00 &  0.22  \\
                        % Materials &   \textcolor{purple}{0.718}   &  0.00 &  0.00 &  0.00 &   0.141\\
                        %Werkstoffpraktikum  &   \textcolor{purple}{0.577}   &  0.00 &  0.00 &  0.00 &   \textcolor{orange}{0.096}\\
                        BusinessAdmin  &  0.00 &  0.00 &  0.00 &  0.00 &  0.21  \\
                        Chemistry  &  0.00 &  0.00 &   \textcolor{purple}{0.09}   &  0.00 &   \textcolor{purple}{0.09}   \\
                        ElectEng  &  0.00 &  0.00 &  0.00 &  0.00 &  0.18  \\
                        ControlEng &  0.00 &  0.00 &  0.00 &  0.00 &  0.15  \\
                        FluidMech  &  0.00 &  0.00 &  0.00 &  0.00 &  0.14  \\
                        ConstructEng I  &  0.00 &  0.00 &  0.00 &  0.00 &  0.12  \\
                        ConstructEng II  &  0.03  &  0.00 &  0.00 &  0.00 &   0.10   \\
                        Mathematics I  &  0.01  &  0.00 &  0.00 &  0.00 &  0.12  \\
                        Mathematics II  &   \textcolor{purple}{0.05}   &  0.00 &   \textcolor{purple}{0.11}   &  0.00 &   \textcolor{purple}{0.03}   \\
                        Mathematics III  &  0.00 &  0.00 &  0.00 &  0.00 &  0.11  \\
                        Mechanics I  &  0.00 &  0.00 &  0.00 &  0.00 &  0.18  \\
                        Mechanics II  &  0.00 &  0.00 &   \textcolor{purple}{0.25}   &  0.00 &   \textcolor{purple}{0.04}   \\
                        NumMath  &  0.02  &  0.00 &  0.00 &  0.00 &  0.13  \\
                        Physics  &  0.00 &  0.00 &  0.00 &  0.00 &  0.28  \\
                        ThermoDyn  &  0.00 &  0.00 &  0.02  &  0.00 &  0.12  \\
                        Materials  &  0.00 &  0.00 &  0.00 &  0.00 &  0.15  \\
                        IndustrialMgmt  &  0.00 &  0.00 &  0.00 &  0.00 &   \textcolor{purple}{0.09}   \\
                        %Werkstoffpraktikum  &  0.00 &  0.00 &   \textcolor{purple}{0.96}   &   \textcolor{purple}{0.49}   &   0.1   \\
                        \bottomrule
                \end{tabular}
                }
          \end{minipage}
    
    \label{tab:mar}
\end{table}
            Since we find relationships between the non-missing grades and the missingness of student grades, we conclude that MAR is likely in both majors. Courses with low pseudo $R^2$ values are outliers and must be interpreted cautiously. We have copied and transformed the datasets so that each dataset is available twice, once with binary grades for IRT and once with continuous grades for AGM and the centering approach. It is sufficient for the MAR condition check to run the tests on the continuous grade datasets in this context since the mechanism remains the same for both grade types. However, separation is essential for imputation, which we must do next to apply PCA to the datasets. 
            
            \paragraph{Principal Component Analyses}
            Following Section \ref{mipca_imputation}, we use MIPCA under Pearson correlation for continuous grades and MIPCA under tetrachoric correlation for binary grades, achieving imputed complete datasets for CompSci and MechEng. Then, we apply PCA, again depending on the grade scale, to the datasets to estimate the amount of variance that can be explained by the first few principal components. Table \ref{tab:selection_results} shows the explained variance for the first two PCs on each dataset. The PCA on the continuous version of each dataset is in the row 'PCA continuous ($n = 2$)' and on binary data in the following row. 
            
            The simulated datasets give us an upper bound of what one can expect for the first PC. For the one-dimensional dataset, the first two PCs represent $81.16\%$ and $1.80\%$ for continuous data and $38.13\%$ and $5.24\%$ for binarized data. Similarly, for the two-dimensional dataset, the first two PCs represent $71.99\%$ and $13.01\%$ for continuous data and $32.29\%$ and $10.95\%$ for binarized data. Thus, binarizing the data does seem to have a decreasing effect on the relative amount of variance explained by PCA, which is in line with tetrachoric correlation research \citep{kolenikov2004use}.

            We also computed the first two PCs in the continuous and binary cases for CompSci and MechEng. For both datasets, we observe in the continuous case that the first PC covers variances less than $63\%$, which is closer to the 2-dimensional simulated dataset. However, the second PC explains less than $8\%$ of the total variance, which is closer to the 1-dimensional simulated data set for CompSci but not for MechEng. Thus, we cannot directly decide between one and two dimensions and must consult the Bayesian Information Criterion (BIC) results, which compare the models of different dimensionality.
            In the binary case, similar to the simulated data, binarization and subsequent correlation estimation using tetrachoric correlation seem to reduce the proportion of variance explained by the first PC. The second PCs represent larger proportions of the total variances of $6.53\%$ for CompSci and $8.00\%$ for MechEng, which are between the values of the one- and two-dimensional simulated data sets. This again shows that we need the BIC as a complementary criterion to decide if the second dimension is worth including.  
  \begin{table*}[t]
    \centering
    \caption{Model selection results. For each studied major (and major pairing) dataset, we first identified the best-fitting IRT model based on the BIC criterion. Afterwards, we verified that the assumptions of the identified IRT model are fulfilled and that the model parameter fit is reliable and valid.}
    \resizebox{\textwidth}{!}{
        \begin{tabular}{c|c|c|c|c|c|c|c|c|c|c|c|c}
        \toprule
             & \multicolumn{3}{c|}{\textbf{CompSci}} & \multicolumn{3}{c|}{\textbf{MechEng}} & \multicolumn{3}{c|}{\textbf{Simulated 1 Dim}} & \multicolumn{3}{c}{\textbf{Simulated 2 Dim}} \\
             
            \midrule
            No. Students & \multicolumn{3}{c|}{664} & \multicolumn{3}{c|}{1651} & \multicolumn{3}{c|}{2000} & \multicolumn{3}{c}{2000} \\
            No. Courses & \multicolumn{3}{c|}{19} & \multicolumn{3}{c|}{18} & \multicolumn{3}{c|}{20} & \multicolumn{3}{c}{20} \\
            No. Course Offerings & \multicolumn{3}{c|}{127} & \multicolumn{3}{c|}{177} & \multicolumn{3}{c|}{} & \multicolumn{3}{c}{} \\
            \hline
             \multicolumn{13}{c}{\textbf{1. Dimensionality}} \\
             \hline
            Little's test 
            & \multicolumn{3}{c|}{not likeli MCAR} 
            & \multicolumn{3}{c|}{not likeli MCAR} 
            & \multicolumn{3}{c|}{no missing vals}  
            & \multicolumn{3}{c}{no missing vals} \\
            Logistic Regression Test 
            & \multicolumn{3}{c|}{likeli MAR} 
            & \multicolumn{3}{c|}{likeli MAR}  
            & \multicolumn{3}{c|}{no missing vals}   
            & \multicolumn{3}{c}{no missing vals} \\
            PCA continuous ($n = 2$) 
            & \multicolumn{3}{c|}{$62.5\%$, $5.6\%$} 
            & \multicolumn{3}{c|}{$48.0\%$, $7.8\%$} 
            & \multicolumn{3}{c|}{$81.16\%$, $1.80\%$} 
            & \multicolumn{3}{c}{$71.99\%$, $13.01\%$} \\
            PCA binary ($n = 2$) 
            & \multicolumn{3}{c|}{$50.13\%$, $6.53\%$} 
            & \multicolumn{3}{c|}{$34.40\%$, $8.00\%$} 
            & \multicolumn{3}{c|}{$38.13\%$, $5.24\%$} 
            & \multicolumn{3}{c}{$32.29\%$, $10.95\%$} \\
            BIC AGM 
            & \multicolumn{3}{c|}{1 Dim} 
            & \multicolumn{3}{c|}{1 Dim} 
            & \multicolumn{3}{c|}{1 Dim}
            & \multicolumn{3}{c}{2 Dim}\\
            BIC IRT 
            & \multicolumn{3}{c|}{Rasch} 
            & \multicolumn{3}{c|}{Rasch} 
            & \multicolumn{3}{c|}{Rasch}
            & \multicolumn{3}{c}{2PL-2Dim}\\
            % Residual PCA IRT & \multicolumn{3}{c|}{$10.69\%$, $9.38\%$} & \multicolumn{3}{c|}{$8.55\%$, $6.22\%$} & \multicolumn{3}{c|}{$10.38\%$, $9.92\%$} & \multicolumn{3}{c}{$0.0\%$, $0.0\%$}\\
            % Residual PCA AGM & \multicolumn{3}{c|}{$5.72\%$, $3.24\%$} & \multicolumn{3}{c|}{$10.13\%$, $7.79\%$} & \multicolumn{3}{c|}{$9.57\%$, $4.81\%$} & \multicolumn{3}{c}{$4.95\%$, $2.74\%$}\\
            % Residual PCA Centering & \multicolumn{3}{c|}{$59.7\%$, $6.2\%$} & \multicolumn{3}{c|}{$47.0\%$, $7.8\%$} & \multicolumn{3}{c|}{$87.57\%$, $10.82\%$} & \multicolumn{3}{c}{$54.62\%$, $28.97\%$} \\
            
            \hline
             \multicolumn{13}{c}{\textbf{2. Local independence}} \\
                          \hline
             & IRT & AGM & Centering 
             & IRT & AGM & Centering 
             & IRT & AGM & Centering
             & IRT & AGM & Centering \\
             \hline
             Centering Q3 
             & $-0.06$&$-0.18$&$-0.14$
             & $-0.06$&$-0.10$ &$-0.10$
             & $-0.09$&$-0.11$&$-0.11$
             & $ 0.37$&$-0.07$& $-0.05$ \\
            Q3 violations 
            & $3/171$ &  $4/171$ & $9/171$ 
            & $1/153$ & $10/153$ & $10/153$ 
            & $0.0/190$ & $0.01/190$ & $0.42/190$ 
            & $46.4/190$ & $18.3/190$ & $123.7/190$\\

            \hline
             \multicolumn{13}{c}{\textbf{3. Time Invariance}} \\
             % \hline
             % Centering Match Test & & & & & & & \\
             \hline
             & IRT & AGM & Centering 
             & IRT & AGM & Centering 
             & IRT & AGM & Centering 
             & IRT & AGM & Centering \\
             \hline
             
            Student Time Split-Half 
             & $0.64$ & $0.75$ & $0.65$
             & $0.48$ & $0.70$ & $0.67$
             & $0.82$ & $0.97$ & $0.97$
             & $0.73$ & $0.96$ & $0.92$\\

            Course Time Split-Half 
             & $0.59$ & $0.75$ & $0.78$
             & $0.80$ & $0.82$ & $0.78$
             & $0.99$ & $0.99$ & $1.00$
             & $0.98$ & $0.96$ & $1.00$\\
             \hline
             \multicolumn{13}{c}{\textbf{Reliability}} \\
             \hline
             & IRT & AGM & Centering
             & IRT & AGM & Centering 
             & IRT & AGM & Centering
             & IRT & AGM & Centering \\
             \hline

            Student Random Split-Half
            & $0.81$ & $0.88$ & $0.87$ 
            & $0.71$ & $0.83$ & $0.82$ 
            & $0.81$ & $0.97$ & $0.97$ 
            & $0.69$ & $0.97$ & $0.92$ \\

            Course Random Split-Half
            & $0.97$ & $0.98$ & $0.97$ 
            & $0.98$ & $0.99$ & $0.99$ 
            & $0.99$ & $1.00$ & $1.00$ 
            & $0.99$ & $0.98$ & $1.00$ \\
            
            \hline
             \multicolumn{13}{c}{\textbf{Validity}} \\
             \hline
             & IRT & AGM & Centering 
             & IRT & AGM & Centering  
             & IRT & AGM & Centering  
             & IRT & AGM & Centering \\
             \hline
            Student Parameters 
            & $0.98$ &$0.99$ & $1.00$ 
            & $0.97$ &$0.99$ & $0.99$ 
            & $0.99$ &$0.99$ & $1.00$
            & $0.90$ & $0.92$ & $1.00$\\
            Course Parameters
            & $0.89$ &$0.84$ &$0.86$
            & $0.95$ &$0.99$ & $0.99$ 
            & $0.99$ &$0.99$ & $1.00$
            & $0.96$ & $0.65$ &$1.00$\\
        \bottomrule

        \end{tabular}
        }
    \label{tab:selection_results}
\end{table*}

            \paragraph{Bayesian Information Criterion}
            Now that PCA tells us how many dimensions come into question, we fit models with the appropriate dimensionalities on the data sets that include missing values. For the centering approach, we can not fit multidimensional models based on the design of the approach. We now compare the fitted IRT and AGM models (1 and 2-dimensional) using BIC (only within their model class). For the simulated datasets, as expected, we get the best fit for models with a dimensionality according to the datasets' ground truth dimensionality in both grade scales, binary and continuous (cf. rows 'BIC AGM' and 'BIC IRT'). The CompSci and MechEng datasets best fit a one-dimensional binary and continuous model. Thus, we continue with one-dimensional IRT and AGM models for the CompSci and MechEng datasets.

        \subsubsection{Assumption 2: Local Independence}
            For the second assumption, local independence, we calculate Yen's Q3 criterion using the residuals between the dataset and the modeled grades. For the binary data sets, we use the predicted pass probabilities of the IRT models instead of the modeled 'pass'/'fail' grades to obtain continuous scales for the residuals and thus be able to compute a Pearson correlation. For each dataset and each model, we show the average Q3 value and the number of course-pair violations in Table \ref{tab:selection_results}. 
            For all one-dimensional data sets, the number of Q3 violations is lowest for the IRT model. The AGM model and the centering approach have the most violations on average. 
            For the two-dimensional simulated data, we see more violations for all models. The centering approach again has the most violations, on average 123.7/190, indicating that it cannot capture the underlying structure of the data well, likely because the second dimension is not modeled explicitly (unlike IRT and AGM). However, the multidimensional IRT (46.4/190) and AGM (18.3/190) models also violate the LI condition more often than in the one-dimensional cases. This may be due to parameter identification problems, well-known for multidimensional IRT models \citep{Ayala2013:Theory}. 
            According to the methodology, we must merge all course pairs that violate the Q3 condition. However, for the sake of intermodel comparability, we choose to model the courses separately and interpret the results cautiously. Thus, we conclude that for the CompSci and MechEng datasets, the IRT, AGM, and Centering Approach models satisfy the LI assumption for the vast majority of course pairs.

        \subsubsection{Assumption 3: Time-Invariance}
            For the third assumption, time invariance, we examine the split-half test, where grades of each student are split into the first and second halves of their university career. The models fitted independently (for each model type) on the halves give us parameter sets to compare. For all models, we obtain sufficiently high correlations for all data sets ($>0.6$), which supports the time-invariance assumption of the models. 
        
            We have shown that the assumptions of the selected IRT and AGM models are mostly fulfilled. The centering approach violates the local independence and time invariance assumptions more often than the latent variable models, suggesting that it may not be flexible enough to capture the underlying structure of the data well. The centering approach performs worse, especially when the underlying structure of the data is multidimensional. 
    
    \subsection{Assessing Validity and Reliability}
        For \textit{validity}, we compare the fitted model parameters against course average grades and student GPAs. All model-dataset combinations show high correlations ($>0.6$) for student and course parameters, indicating that the fitted parameters capture the concepts we intended, i.e., course difficulty and student performance. 
    
        For \textit{Reliability}, we compare the parameter sets resulting from models fitted on random studentwise split-half partitioning. Again, all model-dataset combinations show high correlations ($>0.6$), indicating a robust parameter fit.

        \subsection{Centering Approach Fails Validity In Biased Settings}
         
\begin{figure}[t]
                \centering
                \includegraphics[width=1\linewidth]{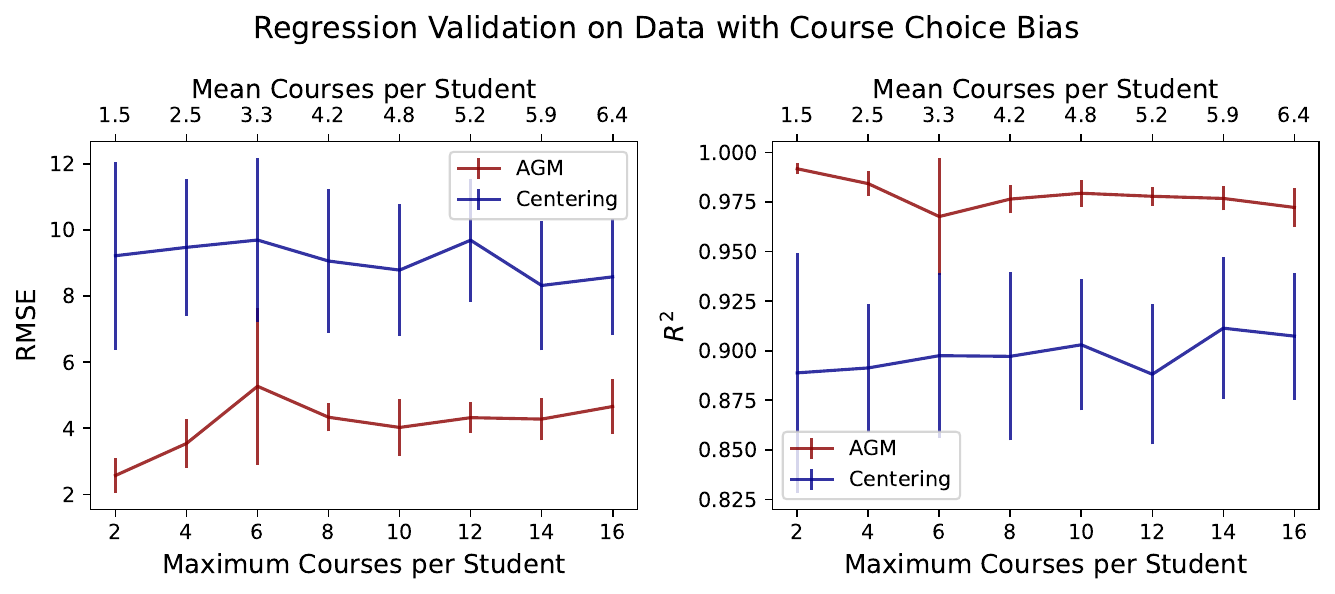}
                \caption{Regression validation results between the centering approach (blue) and AGM (red) on simulated datasets with course choice bias. Over different numbers of courses per student (mean courses per student on the top x-axis and maximum courses per student on the bottom x-axis), the RMSE [left] and $R^2$ [right] show consistently better results for the AGM model, indicating its superior validity over the centering approach estimates.}
                \label{fig:regression_validation}
            \end{figure}

            So far, we have not shown whether latent trait models are advantageous compared to the baseline-centering approaches in terms of assumptions or measurement properties. The simulated data sets used so far are very simple, so the centering approach performs almost equally well. We perform a validation experiment on more unbalanced data in Figure \ref{fig:regression_validation} to show that AGMs have much more stable estimates. To do this, we simulated eight different datasets $X_{sim_1},...,X_{sim_8}$, similar to the one-dimensional simulated dataset, with the addition that students enroll in courses with 10\% chance. Additionally, students with above-average performance traits have a higher chance of enrolling (90\%) in difficult courses, and students with below-average performance traits have a higher chance of enrolling (90\%) in easy courses. Each of the eight datasets $X_{sim_{(\cdot)}}$ has a different maximum number of courses per student to investigate how the validity of the parameters depends on the number of courses per student. The maximum number of courses combined with performance-dependent enrollment results in a mean number of courses per student that is less than this maximum. We then fitted a centering model and an AGM to each dataset $X_{sim_{(\cdot)}}$. After model fit, we were left with student and course estimates for both models on each dataset. For each of the simulated datasets $X_{sim_{(\cdot)}}$, we created two new datasets from the student and course estimates of the two models, one per model. Each row in the two datasets consists of the respective model's student and course estimate and the corresponding course grade (e.g., for the AGM, a row is similar to ($\phi_s$, $\delta_c$, $(X_{sim_{(\cdot)}})_{c,s}$). This results in 16 datasets. These datasets were split with train-test splits (70/30) and fitted with linear regression models. The results are presented in terms of root mean squared error (RMSE) and $R^2$. The process is repeated 10 times to generate confidence intervals. We can see in Figure~\ref{fig:regression_validation} that the latent AGM performs significantly better in terms of RMSE and $R^2$ and has smaller confidence intervals. This highlights how latent models are more robust to biases, such as a course choice bias, and should be preferred over centering approaches.  
           
    \subsection{Generate Insights}
        Having provided the central purpose of the paper, providing an accessible tutorial and the CDE package for novel CA methodologies that enable researchers and practitioners to address various questions of interest, we now illustrate the utility of the generated difficulty measures for answering multiple potential research questions (outlined in Table \ref{tab:research_questions}).

        \subsubsection{Do external events influence the course difficulties at my university?}\label{sec:monitoring_difficulty}
\begin{figure}[t]
                \centering
                \includegraphics[width=1\linewidth]{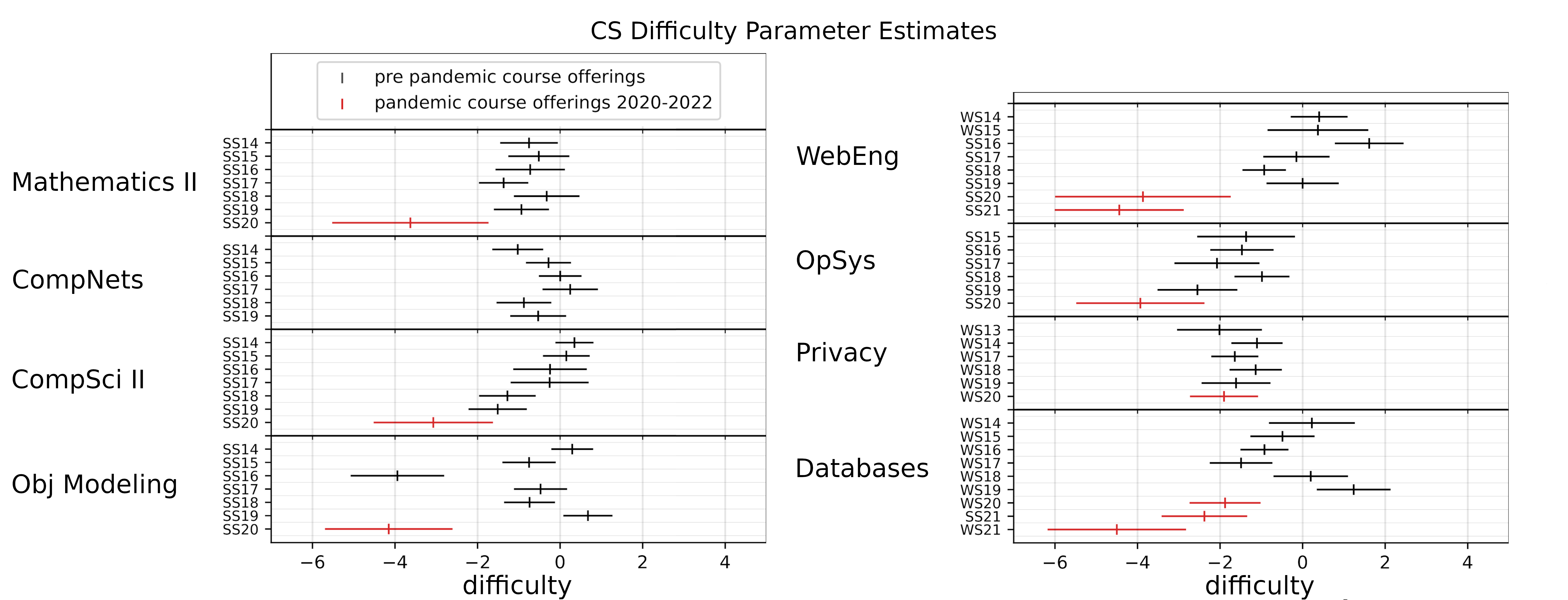}
                \caption{Difficulty of computer science courses over time determined by the Rasch model modeling individual course offerings. The $95\%$ confidence intervals are determined using bootstrapping. The offerings of individual courses show different developments over time (stationary, in-/decreasing). In particular, the red offerings are offerings during the COVID-19 pandemic and show a systematic downwards shift in course difficulty.}
                \label{fig:course_diff_over_time}
            \end{figure}

            To monitor the evolution of course difficulty over time we fit one parameter per course offering rather than one parameter per course \cite{baucks24lak}. For example, if a course was offered six times in three years, we fit a course difficulty parameter six times. Here, we use IRT on binary data ('pass'/'fail' grades). In Figure \ref{fig:course_diff_over_time}, we can see that course difficulty can change over time. In particular, the courses marked in red are course offerings during the COVID-19 pandemic. For these, we were able to identify a statistically significant drop that was not known to the university stakeholders before. This observation opens avenues for future studies to explore the factors that drive changes in course difficulty over time, such as changes in instructional practices, assessment strategies, or institutional resources, and their broader implications for promoting fairness and equity.

        \subsubsection{Do courses exhibit implicit biases that impact groups disproportionately?}\label{sec:dcf_result}
            As an extension of the latent models, we discussed DCF detection, which is based on the idea of Differential Item Functioning \citep{Baucks24Gaining_LAS}. DCF allows the quantification of group-specific differences in experienced course difficulty. Depending on the features apparent in the data, we can divide the students into groups to see whether students in one group find individual courses more difficult, independently of the students' individual performance traits and the course's difficulty estimate. Here, we illustrate DCF analyses by partitioning students into dropout/graduation groups and well as groups beginning their studies before/after $2016$.

\begin{table}[t]
            \centering
            \caption{Significant DCF effects between drop-outs (positives are easier) and graduates (negatives are easier) after Benjamini-Hochberg (BH) correction.
                        All DCF results in the tables show BH-$p$-value $<0.05$}
            \resizebox{0.49\textwidth}{!}{
            \begin{tabular}{cccccc}
                \toprule
                \textbf{Course} & \textbf{Dropouts} & \textbf{Graduates} & \textbf{DCF}\\
          
                \hline
                 \hline
                \multicolumn{4}{c}{\textbf{CompSci IRT}} \\
                \hline
                CompNets &  225 & 203 & -0.621 \\
                CompArch &  77 & 205 & -0.575 \\
                WebEng &  60 & 206 & -0.539 \\

                \hline
                \multicolumn{4}{c}{\textbf{CompSci AGM}} \\
                \hline
                Mathematics II & 206 & 196 & -25.256 \\
                Obj Modeling & 222 & 205 & -22.547 \\
                CompSci II   & 218 & 204 & -20.968 \\
                Management   & 171 & 203 & -20.523 \\
                CompNets   & 225 & 203 & -19.729 \\
                Mathematics I   & 229 & 196 & -17.972 \\
                DiscMath &   95 & 205 & -17.39 \\
                CompSci I &   236 & 205 & -16.953 \\
                CompArch &   77 & 205 & -16.914 \\
                CompSci III &  99 & 204 & -16.571 \\
                Programming &   228 & 202 & -16.222 \\
                Statistics &   234 & 204 & -16.152 \\
                WebEng &   60 & 206 & -16.049 \\
                Databases &   60 & 206 & -15.784 \\
                OpSys &   62 & 203 & -15.193 \\
                Daten Structures &  55 & 205 & -14.737 \\
                Economics &   237 & 204 & -13.663\\
                SoftEng &   63 & 204 & -7.257\\
                Privacy &   126 & 205 & -6.892 \\
                \bottomrule
            \end{tabular}
            }
            \hspace{0.01cm}
            \resizebox{0.49\textwidth}{!}{
                \begin{tabular}{cccccc}
                \toprule
                \textbf{Course} & \textbf{Dropouts} & \textbf{Graduates} & \textbf{DCF} \\                
                \hline
                \hline
                \multicolumn{4}{c}{\textbf{MechEng IRT}} \\
                \hline

                IndustrialMgmt &  104 & 711 & -0.747 \\
                FluidMech &  111 & 716 & -0.56 \\
                ControlEng &   110 & 718 & -0.35 \\
                ThermoDyn &   128 & 725 & -0.321 \\
                ConstructEng I   & 150 & 681 & -0.307  \\
                Physics &   180 & 721 & 0.425 \\
                Mechanics I   & 176 & 677 & 0.578 \\
                Chemistry   & 184 & 724 & 0.582 \\
                
                \hline
                \multicolumn{4}{c}{\textbf{MechEng AGM}} \\
                \hline
                IndustrialMgmt   & 104 & 711 & -17.51  \\
                ElectEng &   150 & 717 & -12.511  \\
                ConstructEng I   & 150 & 681 & -11.809 \\
                NumMath &  160 & 712 & -10.958\\
                ThermoDyn   & 128 & 725 & -10.78 \\
                Materials   & 151 & 672 & -10.617  \\
                Mathematics III   & 142 & 705 & -10.253\\
                FluidMech &   111 & 716 & -10.152 \\
                Mechanics II   & 153 & 709 & -10.056 \\
                Mathematics II   & 169 & 671 & -9.933 \\
                Mathematics I   & 170 & 654 & -9.438 \\
                ControlEng &   110 & 718 & -8.693  \\
                BusinessAdmin   & 128 & 728 & -7.887 \\
                Physics &   180 & 721 & -6.409  \\
                ConstructEng II  & 73 & 694 & -6.271 \\
                %MechEng & AGM & Werkstoffpraktikum &  & 38 & 90 & -4.745 & 0.009 \\
                Chemistry &   184 & 724 & -3.869  \\
                Mechanics I   & 176 & 677 & -3.754 \\
               \bottomrule
            \end{tabular}
            }

            \label{tab:dcf_drop}

        \end{table}

\begin{table}[t]
            \centering
            \caption{Significant DCF effects between cohorts before 2016 (negatives are easier) and after 2016 (positives are easier) after Benjamini-Hochberg (BH) correction. All DCF results in the tables show BH-$p$-value $<0.05$}
            \resizebox{0.49\textwidth}{!}{
            \begin{tabular}{cccc}
                \toprule
                \textbf{Course} &  \textbf{After 16/15 } & \textbf{Before 16/15} & \textbf{DCF} \\
                \hline
                \hline
                \multicolumn{4}{c}{\textbf{CompSci IRT}} \\
                \hline
                CompSci III &  220 & 234 & -0.414 \\
                CompNets & 330 & 314 & -0.364 \\
                Management &  330 & 259 & 0.269 \\
                CompSci II &  332 & 311 & 0.345 \\
                WebEng &  217 & 200 & 0.382 \\
                Databases  & 184 & 205 & 0.567  \\

               \bottomrule
            \end{tabular}
            }

            \label{tab:dcf_cohorts}

        \end{table}
        
            % Why are we looking at dropouts?
            The latent trait models assess the traits of students' performance and course difficulty as time-invariant and locally independent. In reality, however, we expect the order in which students take courses or the time interval between courses to play a role (e.g., \citealt{gutenbrunner2021measuring,weiss2022impact} detect differences using centering approaches). DCF can not only detect differences but also quantify them. In particular, it is interesting to investigate dropouts because they tend to fail courses and, therefore, deviate from the program's recommended course sequence or take longer breaks between courses. We expect dropouts to have different prior knowledge than graduates when enrolling in courses. The student performance trait does not account for that. 
        
            % What are the results?
            Table \ref{tab:dcf_drop} shows significant DCF effects between dropouts and graduates for both majors, CompSci and MechEng, in both models, IRT and AGM. To adjust the significance level for multiple testing, we adjusted the false discovery rate (FDR) of each test using the Benjamini Hochberg (BH) correction with the target FDR value of $0.05$ \citep{benjamini1995controlling}. Here, positive effects mean that dropout students found the course easier, and negative effects mean that graduating students found the course easier. Looking at the difference between binary and continuous modeling, we detect more significant DCF effects for AGMs. This is expected since AGMs can more precisely model information in the 'pass' bin than IRT models.
            %, e.g. if a difference between groups exists only in the 'pass' bin. For CompSci and MechEng, this results in more significant effects being detected when the entire grade scale is used. 
            Still, IRT's effects of passing a course stay valid. However, the ranking of significant DCF effects changes for some courses compared to DCF detection on AGMs. This shows that IRT-related DCF effects are not generalizable to the whole grade scale.  
        
            For AGM, courses that do show significant DCF effects and, in addition, have a consecutive course (e.g., Mathematics I - II in CompSci or Mathematics I - II - III and Mechanics I - II in MechEng), then that consecutive course shows a significant DCF effect too. That effect is often greater than in the first course. Thus, DCF is often inherited for courses with consecutive content. For IRT in the CompSci major, the effects align with the AGM DCF effects and follow the same order. However, for MechEng, this is not the case—the ordering changes. We even detect DCF effects with opposite signs between IRT and AGM-related DCF ('Physics,' 'Mechanics', and 'Chemistry'). These courses are easier for dropouts to pass but still more difficult to achieve good grades on the continuous grade scale.  
            Comparing CompSci and MechEng, we observe less significant DCF effects for CompSci under IRT than MechEng. This indicates that dropout students experience courses in MechEng more often as more difficult to pass than in CompSci. However, comparing the AGM DCF effects on the continuous grade scale, we detect not only more DCF effects for CompSci than MechEng but, in addition, larger effects. This indicates that the difference between dropout and graduate students in the CompSci courses is larger on average than in MechEng courses. The identified differences in DCF patterns across programs prompt further investigations of factors causing these inequities, including curriculum structure, grading practices, and differences in student preparation or support systems.
        
        \subsubsection{Is the degree fair for students from different cohorts? }
            For cohorts, we modeled course difficulty for each course \textit{once}, which cannot capture changes that occur within a course over time but instead models the average course difficulty of the courses students in a cohort participated in, as our simulations show in Figure \ref{fig:invariance_assumption}. 
            Splitting the students according to the median starting date of their studies (2016), we can detect a significant change in course difficulty between cohorts in 6/19 CompSci courses in Table~\ref{tab:dcf_cohorts}. In the cohorts that started their studies after 2016, most courses (4/6) for which a DCF effect is detected are perceived as easier. This finding correlates with the systematic decrease in course difficulty that students experienced during the COVID-19 pandemic (see Figure~\ref{fig:course_diff_over_time}). To detect the effect of the COVID-19 pandemic, we separated courses by semester. However, seperating students into cohorts is less nuanced. For example, students in a cohort may take courses in all semesters. This has to be taken into account when comparing the two results. For example, the two courses 'CompSci III' and 'CompNets' with negative DCF values were perceived as more easy by the earlier cohorts before 2016, which is not apparent from the development of course difficulty (see Figure~\ref{fig:course_diff_over_time}). There, for example, the course 'CompNets' is relatively stable over time and the course 'CompSci III' fluctuates. There are various possible reasons for this. Firstly, students from different cohorts may be more evenly distributed over time in that particular course, averaging out the time-dependent difficulty; secondly, the size of courses may fluctuate over time, giving more weight to earlier courses than later courses.
            Overall, this highlights the ability of DCF to capture aspects of the fitted parameters that need further investigation. Some aspects that are not captured by the model parameters of both the IRT model and AGM can be detected and quantified using DCF.
            The findings underscore the importance of examining cohort-level factors -- such as enrollment patterns, class sizes, and semester schedules -- to better understand their impact on the differences in course difficulty as captured by the DCF.

\section{Discussion}\label{sec:discussion}
    This paper presents a comprehensive recipe, in particular, for estimating course difficulty within curriculum analytics (CA), including a GPA-based centering approach and latent variable models based on item response theory (IRT) and additive linear models (AGM). Our aim is to empower CA researchers and practitioners to answer their course difficulty-related questions. Ensuring statistical validity, reliability, and applicability of course difficulty models in educational settings is important but complex. Our tutorial and the open-source 'Course Difficulty Estimation' (CDE) package address the underlying assumptions and methodological challenges aiming to make these advanced techniques accessible to researchers and practitioners. We showcase the utility of the analysis framework based on example datasets from a German university and two simulated datasets. The findings suggest that the latent models can extract valuable insights based on course grade data from higher education institutions.

    We find evidence from model assumption checking guidelines that latent variable models are more flexible and allow quantifiable group analyses or even to adjust multidimensional contexts \citep{Baucks24Gaining_LAS}. In addition, if there are biases in the data, e.g., trait-dependent course choices, the centering approach cannot control for them. The latent models, however, can control for systematic biases in the data, as the regression validation experiment shows.
    
     % DCF to handle methods limitations
    Both model variants, IRT and AGM, fit constant parameters. This means that a student has the same parameter in all courses, and courses have the same parameter for all students. It is possible that students from different groups, e.g., dropouts/graduates or transfer/native students, may have different group-specific course difficulties in individual courses.
    We consider differential course functioning (DCF) detection~\cite{baucks24lak} as a methodological extension of both latent models (i.e., IRT and AGM). DCF can assess course-specific difficulty factors related to students' attributes by analyzing grades from different subgroups. This is useful to detect and quantify differences between groups, e.g., unintended differences between native and non-native speakers. 
    With DCF detection, we can measure group-specific differences in course difficulty independent of students' fitted trait values. This allows us to generate interesting statistics that are of utility in addressing fairness-related questions, e.g., do transfer students find courses more difficult? The detected DCF effect can promote equity by allowing for group-specific support, e.g., do transfer students need additional preparation courses? It also allows for the detection and quantification of violations of the model assumptions of local independence and time-invariance of parameters.

    Using the datasets from the German university as examples, we have generated insights that underscore the utility of implemented models and analyis pipeline. Firstly, detecting changes in course difficulty over time allows stakeholders to monitor difficulty retrospectively, independent of the performance of participating students. The example on the CompSci dataset demonstrates this by flagging a systematic decrease in difficulty during the COVID-19 pandemic that was previously unknown to faculty stakeholders. Secondly, for both data sets, CompSci and MechEng, we found significant DCF effects between dropouts and graduates, indicating that courses are more difficult for dropouts to achieve high grades for but not always more difficult to achieve a pass. Regarding relaxing the model assumptions, DCF effects mainly increase in consecutive courses (e.g., Mathematics I and II in both majors), indicating a potential dependence that may violate the local independence assumption but that can be captured by DCF detection. In addition, we calculated DCF effects between cohorts before and after the median entering semester (2016). Again, we found significant DCF effects for 6/19 courses. This indicates the existence of variation in course difficulty over time, which we have already shown using time-dependent course modeling. Thus, cohort-related DCF further highlights that DCF can quantify and mitigate potential assumption violations of the conventional IRT and AFM models.  
   
    \subsection{Further Considerations and Extensions}
        \begin{itemize}[leftmargin=*]
            \item \textit{What about other applications?} In Figure \ref{fig:method}, we highlight potential insights generateable from the recipe. In detail, we have presented results on difficulty monitoring and DCF. However, the difficulty estimates are applicable across a variety of CA contexts (Table~\ref{tab:research_questions}). Firstly, dashboards are important for reporting results from complex statistical analyses to stakeholders such as student advisors and curriculum policymakers \citep{baucks2024book}. Student advisors can use potential multidimensional course difficulty estimates to propose courses for students to attend, e.g., preparatory courses. Policymakers can monitor difficulty and, where appropriate, introduce closer inspection when courses show significant changes over time. Secondly, student flow models and simulations can benefit from robust difficulty estimates and student performance estimates. These methods are used to understand students' movement through the curriculum better and make predictions about the impact of potential changes to the curriculum (e.g., \citealt{slim2014employing,molontay2020characterizing,saltzman2012simulating}). Robust course difficulty and student performance estimates can make those analyses more reliable. Finally, many other applications can benefit from the estimates because they are fundamental statistics. For example, articulation problems can be extended to include the course difficulty aspect to construct fairer articulation pairs between institutions \citep{pardos2020university}, i.e., are courses not only related in content but equally difficult? 
            \item \textit{How much data do we need?} The models we have investigated have the advantage of being very data efficient compared to deep learning approaches such as autoencoders and advanced probabilistic models such as Markov and Bayesian networks \citep{slim2014employing}. Related research has also done simulation studies to determine how much data is needed or how many missing values are acceptable. These refer mainly to IRT models, but their complexity is similar to AGMs. A good fit can be expected for small data sets of $\geq75$ students per course \citep{baucks24lak}. In addition, there should be at least $10\%$ observed values per course \citep{haas2023bayesian}. However, our simulation of the reliability of imputation-based dimensionality assessments (see Figure \ref{fig:impuation_reliability}) shows that a observed value ratio less than $60\%$ can lead to an underestimation of the underlying structure of the data. Future work can explore how missing data affects difficulty models, from assumptions to outcomes to interpretations in different CA settings. 
            \item \textit{How to use categorical data as it is?} Based on the foundations presented in this paper, further methodological refinements can be made to the recipe. In Section \ref{grade_scale}, we elaborate on grade scales and respective model choices. For example, we model categorical grades as binary or rescale them to a continuous ratio grade scale. However, in practice, we emphasize the importance of keeping as much information as possible. Thus, we could also keep the categorical grades and instead use models suited explicitly for this \citep{veas2017comparative}. For the dimensionality-related PCA, similar to the binary case, we can use polychoric correlation \citep{kolenikov2004use} instead of tetrachoric correlation. For the models, we would then have to fit, for example, a graded response model or a partial credit model that fits an item response function (IRF) for each grade category, similar to the single IRF of the IRT model \citep{Ayala2013:Theory}. In estimation, it gets more complicated because one has to decide how to compute a one-dimensional course difficulty, since categorical models result in one course difficulty per grade category (e.g., \citealt{ali2015location}). 
            \item \textit{How to model temporal variations in student ability?}  One should be careful when interpreting student performance trait values as the "ability to achieve a certain grade (e.g., pass for IRT) in courses on the first attempt" as they might be more constant than more fine-granular aspects of student knowledge. Assessing the models' time invariance assumptions on student and course parameters, we were able to show that the mean is fitted even when time dependency drift exists (e.g., a constant learning rate of students). From the perspective of student parameters, this confirms our split-half experiments, where the mean of the first half of students' courses is compared to the second half. Here, we show that the student parameters are strongly correlated. This is consistent with latent models that fit student learning rates and show very constant small rates \citep{koedinger2023astonishing}. Therefore, time-invariant student parameters are a simplifying assumption that, if violated, does not strongly falsify the interpretation because growth rates appear to be mostly constant. On the other side, for courses, this might not be the case. Here, sudden discontinuities (e.g., the COVID-19 pandemic \citep{baucks24lak}) can occur over time. By time-dependent modeling of the CompSci major, we show that there can be significant fluctuations in the course parameters and, particularly, a systematic drop in difficulty during the pandemic. If ignoring the temporal resolution, our simulation shows that the mean over time is fitted, even when a sudden jump occurs, similar to the student parameters. Time-invariant modeling may be sufficient to compare courses globally as long as one knows that the mean is fitted and interprets the parameters accordingly (i.e., avoid making statements about course offerings in individual semesters). Thus, if enough data is available, and the assumptions can be checked, it makes intuitively sense to consider the courses time-resolved. The mean over time in a time-invariant model can still be calculated from the time-varying estimates.
        \end{itemize}

    \section{Conclusion}
        In summary, this tutorial presented a recipe for estimating course difficulty under different data types using probabilistic latent models (i.e., item response theory and additive grade point models) and heuristic approaches (i.e., centering). We introduced an analysis pipeline for researchers and practitioners, making a `Course Difficulty Estimation` (CDE) package openly available to ensure the rigorous and correct application of these complex statistical methodologies. The procedure yields reliable and valid course difficulty estimates that can be used to address various curriculum analytics questions. Our experiments on data from two undergraduate programs (CompSci and MechEng) demonstrate the utility of latent probabilistic course difficulty models to disentangle course difficulty from student performance. Additional experiments on simulated datasets demonstrate the advantages of these methodological improvements, showcasing their ability to address this limitation inherent in heuristic estimation approaches. Presented extensions of the methods, such as Differential Course Functioning (DCF), provide insights into group differences and course difficulty over time. Our work lays solid foundations for future research in quantitative curriculum analytics, e.g., providing better and more formative feedback to students and understanding the quality of courses in a curriculum. 
        
        To encourage researchers to apply our recipe, we are making the CDE package for the experiments available on GitHub. In addition, the two simulated datasets are available for a quick start. To increase the usability of our CDE package, we proposed a standardized course response format as shown in Section \ref{sec:hands_on}, which makes the application straightforward. We hope this repository will benefit future CA research and make these complex statistical methodologies accessible to a wide community of CA researchers and practitioners to estimate course difficulties easily.

\bibliographystyle{acmtrans}
\bibliography{./ref}

\newpage

\appendix
\section{Little's test for MCAR} \label{app:little}
    
        Little's test \citep{little1988test} is a statistical test to check whether missing values are MCAR, i.e., independent of the observed and unobserved values. The idea behind the test is to calculate a chi-square statistic that measures the deviations between observed and expected means of missing values. 
        
        Suppose we have a course-response matrix $X\in \mathbb{R}^{S \times C}$. There are missing values in this matrix. Assume that the grades in $X$ are multivariate normally distributed by the student. If we knew the missing values, then the grades $x_i \in \mathbb{R}^C$ for each student $i \in (1, ..., S)$ would come from a $C$ dimensional multivariate normal distribution $N(\mu, \Sigma)$.  
        The missingness under MCAR does not depend on the observed or missing grades. This means that if the assumption is correct, we should not be able to find patterns in the missingness. To check this, we assume the opposite, that the missing values are described by patterns $p$ from a set of patterns $P$. A pattern $p \in P$ is defined by two index sets $O_p$ and $M_p$ with $O_p \cup M_p = \{1,...,C\}$, where $O_p$ indicates the observed courses and $M_p$ the missing courses for each student that is part of the pattern $p$. For a given pattern, we can then calculate the mean values $\mu_{O_p}\in \mathbb{R}^{|O_p|}$ and covariances $\Sigma_{O_p}\in \mathbb{R}^{|O_p| \times |O_p|}$ of the observed courses from the assumed ground truth distribution $\mu$ and $\Sigma$. The idea behind Little's test is now to calculate the discrepancy between the expected means $\mu_{O_p}$ under the MCAR assumption and the empirically observed means of the data. To do this, let $\hat{\mu}_{O_p}$ be the empirical observed mean on the observed courses. Further, let $s_p$ be the number of students in pattern $p$, where $\sum_p s_p = S$. Then we calculate the discrepancy overall patterns $p\in P$ as:
        \begin{align*}
          T^2 = \sum_{p\in P} s_p (\boldsymbol{\hat\mu}_{O_p} - \boldsymbol{\mu}_{O_p})^T \Sigma_{O_p}^{-1} (\boldsymbol{\hat\mu}_{O_p} - \boldsymbol{\mu}_{O_p}). 
        \end{align*}
        Then Little \cite{little1988test} has shown that $T^2$ follows a chi-squared distribution with~$n = (\sum_p |O_p|) - C > 0$ degrees of freedom. 
        The null hypothesis $H_0$ then states that the means $\mu_{O_p}$ do not change between the patterns, while the alternative hypothesis $H_1$ states that a separate mean exists for each missing value pattern. In reality, we do not know the ground truth distribution $N(\mu,\Sigma)$ and must, therefore, approximate it using maximum likelihood approaches or take the means and variances of the data set in a simplified way, such as in many implementations \cite{schouten2021dance}. 
        The p-value is the probability of obtaining a test statistic $T^2$ at least as extreme as the one observed, given the null hypothesis $H_0$ is true. This can be found using the cumulative distribution function of the chi-square distribution:
        \begin{align*}
           \text{\textit{p}-value} = P( T^2 | H_0 ) = P(\mathcal{X}_n^2 > T^2) = 1 - P(\mathcal{X}_n^2 \leq T^2) 
           = 1 - \frac{\int_0^{T^2}t^{\frac{n}{2}-1} e^{-t}dt}{(\frac{n}{2}-1)!},
        \end{align*}
        A p-value of less than $0.05$ indicates that means dependent on the patterns are likely to exist, which suggests that the data is likely not MCAR.

\section{Reliability of Explained Variance under Imputation} \label{app:reli_amputation}

 Since PCA arguments were designed for a complete dataset, we want to assess the influence of the number of missing values on the whole dimensionality argument. Thus, we ensure a robust model selection. For that, we assess the reliability of the dimensionality assessment under different rates of missing data that are MAR. 
    We want to check whether the imputation distorts the explained variance of the PCA. This could bias our dimensionality analysis. To do this, we simulate complete datasets and then \textit{ampute} data \cite{schouten2021dance} under the MAR condition. 
    
    Following the pseudocode in Algorithm \ref{alg:amputation}, we assume that students with decreasing trait levels and courses with increasing difficulty have an increasing probability of missing a course grade. This simulates dropping out. For each performance threshold $\tau$, we end up with $10$ different simulation settings. In each setting, we have increasing amputation rates $\alpha \in {0.1, ..., 0.9}$. These describe the probability of missing grades if a course has pass rates or a student has a GPA lower than $\tau$. If grades correspond to courses or students above $\tau$, we assume a missing probability of $0.1$. 
    After amputation, we impute the missing values using MIPCA (typically used for MAR imputation) and mean imputation (typically used for MCAR imputation).
    We then compute PCs using PCA on both the imputed and ground truth complete datasets and compare the variance explained by the PCs on the datasets. If the imputation is reliable, the explained variance of the PCs under missing data remains approximately constant. 

\begin{algorithm}[t]
    \caption{Simulation Study: Amputation under MAR}
    \label{alg:amputation}
    \begin{algorithmic}
    \Require Performance threshold $\tau$, amputation rate $\alpha \in [0,1]$, base rate $\beta = 0.1$, number of simulations $n$.
    \Ensure $n$ datasets with MAR missing values.
    
    \For{$i = 1 \to n$}
        \State \textbf{Simulate complete dataset:}
        \State  \hspace{0.5em}Draw groundtruth Gaussian-distributed student and course parameters $\theta_s$ and $delta_c$
        \State  \hspace{0.5em}Generate course response matrix $X^{(i)} = (g_{s,c})_{s,c}$ using an IRT model 
        \State  \hspace{0.5em}Compute PCA on the course response matrix
        \State  \hspace{0.5em}Compute ground truth explained variance using the first principal component.
        
        \State \textbf{Amputation process:}
        \State  \hspace{0.5em}Calculate GPA for students $\mu_s$ and course pass rates $\mu_c$.
        \State  \hspace{0.5em}\textbf{For each student-course pair $(s, c)$:}
        \State \hspace{1em} \textbf{if} $\mu_s < \tau$ \textbf{or} $\mu_c < \tau$:
        \State \hspace{2em} Set probability of missing grade $P(g_{s,c} \text{ missing}) = \alpha$
        \State \hspace{1em} \textbf{else:}
        \State \hspace{2em} Set probability of missing grade $P(g_{s,c} \text{ missing}) = \beta$
        \State \hspace{0.5em} Apply amputation to the dataset.

        \State \textbf{Impute missing data:}
        \State \hspace{0.5em} Perform mean imputation and MIPCA (Multiple Imputation by PCA).

        \State \textbf{PCA Analysis:}
        \State \hspace{0.5em} Compute principal components and explained variance for imputed datasets.
    \EndFor
    
    \end{algorithmic}
\end{algorithm}

    \begin{figure}[t]
        \centering
        \includegraphics[width=0.83\linewidth]{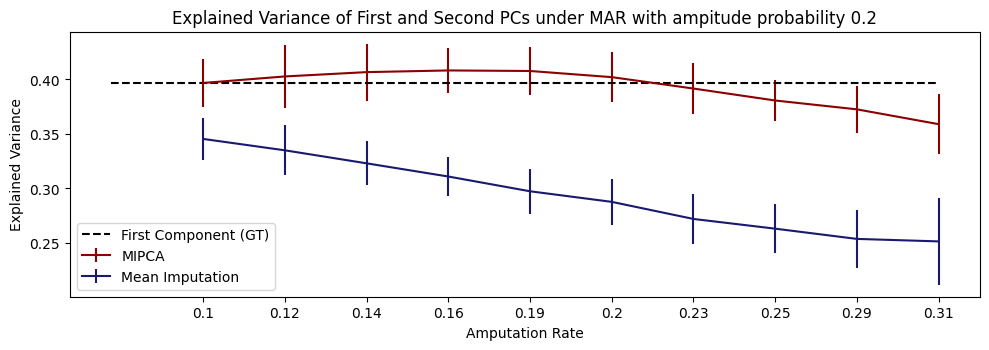}
        \includegraphics[width=0.83\linewidth]{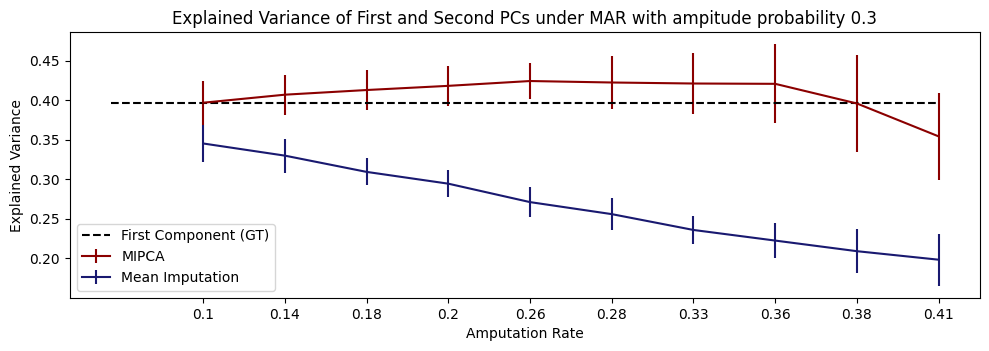}

        \caption{Simulation of the effect of different imputation methods on the variance explained by PCA, highlighting the importance of choosing the right imputation method under the missing at random (MAR) mechanism for reliable imputation. We simulate data with different rates of missing values under the MAR assumption. Missing values are imputed using both mean imputation and multiple imputation. The explained variance of the first principal component is compared to the ground truth. MIPCA imputation closely approximates the true proportion of variance, while mean imputation, which should be applied under the MCAR assumption, significantly underestimates the variance of the first principal component.}
        \label{fig:impuation_reliability}
    \end{figure}
    
    Figure \ref{fig:impuation_reliability} shows simulation results for two performance thresholds $\tau \in \{0.2, 0.3\}$. As the amputation rate increases, the amount of missing data increases. 
    The dotted line represents the ground truth explained variance by the first PC on the complete dataset. For increasing $\alpha$ values, we obtain increasing missing value rates dependent on $\tau$. These range from $0.1$ to $0.41$, which relates to the datasets we will introduce later. 
    The dark blue and dark red lines represent the explained variance of the first PC on the imputed datasets by mean imputation and MIPCA, respectively. The explained variance for mean imputation decreases as the missing rate increases. MIPCA, on the other hand, shows a more stable explained variance close to the ground truth, demonstrating the power of MIPCA to capture the underlying structure of missingness in the data. This underscores the importance of handling missing values under the correct assumption.

\section{Likelihoods for IRT and AGM}\label{app:likelihoods}

In the context of IRT, each course grade $X_{s,c}$ can be modeled as a Bernoulli-distributed random variable. Let $p_{s,c} = P(X_{s,c} =1 \,|\, \boldsymbol{\theta}_s, \boldsymbol{\alpha}_c, \boldsymbol{\delta}_c)$, then the log-likelihood for the IRT models can be written as:
        \begin{align*}
            \hat{\mathcal{L}}_{\text{IRT}}(\boldsymbol{\theta}, \boldsymbol{\alpha}, \boldsymbol{\delta}) 
            %&= log\left( \prod_{s,c} p_{s,c}^{X_{s,c}}(1-p_{s,c})^{1-X_{s,c}} \right)\\
            &= \sum_{s, c} \left(X_{s,c} \log (p_{s,c})
            + (1 - X_{s,c}) \log (1 - p_{s,c} )\right)
        \end{align*}
        For AGM models, we have residuals for each observed value:
        \begin{align*}
            R_{s,c} = X_{s,c} - \sum_{d=1}^D (\boldsymbol{\theta}_s)_{d} + (\boldsymbol{\delta}_c)_{d}.
        \end{align*}
        These residuals can be assumed to be normally distributed. Then the empirical variance of this normal distribution is $\sigma^2 = (\#S\#C)^{-1} \sum_{s,c} R_{s,c}^2$ leading to the log-likelihood to be:
        \begin{align*}
            \hat{\mathcal{L}}_{\text{AGM}}(\boldsymbol{\theta}, \boldsymbol{\delta}) 
            &=\sum_{s,c} log\left[ \frac{1}{\sqrt{2 \pi \sigma^2} }\text{exp}\left(\frac{-R_{s,c}}{2\sigma^2}\right)
 \right] \\
            &= \frac{\#S\#C}{2}\left[\text{log}(2\pi) +  \text{log}\left(\frac{1}{\#S\#C} \sum_{s,c}R_{s,c}^2\right)\right].
            % \\
            % &- \lambda \left( \|\boldsymbol{\theta}_s\|^2 + \|\boldsymbol{\delta}_c\|^2 \right) \\
            % &- ( \|\boldsymbol{\theta}_s^T \boldsymbol{\theta}_s - I\|^2 + \|\boldsymbol{\delta}_c^T \boldsymbol{\delta}_c - I\|^2 ).
        \end{align*}

\end{document}